\documentclass[journal]{IEEEtran}
\pdfoutput=1
\usepackage{ifpdf}
\usepackage{cite}
\usepackage[cmex10]{amsmath}
\usepackage{amsfonts,amssymb}%
\usepackage{graphicx}
\usepackage{color}
\usepackage[tight,footnotesize]{subfigure}
\usepackage{epsf}
\usepackage{eucal}
\usepackage[ruled]{algorithm2e}
\usepackage{array}
\usepackage{url}

\newcommand{\eqdef}{\stackrel{\Delta}{=}}
\newcommand{\eqsp}{\, }

\newtheorem{prop}{Proposition}
\newtheorem{lem}{Lemma}
\newtheorem{theo}{Theorem}

\newcommand{\be}{\beta}
\newcommand{\bbe}{\boldsymbol{\beta}}
\newcommand{\wbbe}{\widehat{\boldsymbol{\beta}}}
\newcommand{\bdelta}{\boldsymbol{\delta}}
\newcommand{\beps}{\boldsymbol{\varepsilon}}
\newcommand{\bzeta}{(\boldsymbol{\delta}+\boldsymbol{\varepsilon})}
\newcommand{\bg}{\mathbf{G}}
\newcommand{\po}{P_{0}}
\newcommand{\ab}{\textbf{A}}
\newcommand{\rset}{\mathbb{R}}

\newcommand{\1}{\mathbf{1}}
\newcommand{\argmin}{\arg\,\min}

\newenvironment{pv}{\textit{Proof}: \hspace{0.1cm}}{$\rule{2.5mm}{2.5mm}$}

\usepackage{hyperref}

\begin{document}
%
\title{Sparse regression algorithm for activity estimation in $\gamma $ spectrometry}
%
%
%

\author{Y.~Sepulcre, T.~Trigano,~\IEEEmembership{Member,~IEEE,}  and~Y.~Ritov
\thanks{Y. Sepulcre is with the Department of Computer Science, Jerusalem College of Engineering, Israel}
\thanks{T. Trigano is with the Department
of Electrical Engineering, Shamoon College of Engineering, Israel. e-mail: thomast@sce.ac.il}
\thanks{Y. Ritov is with the Department of Statistics, Hebrew University of Jerusalem, Israel. email: yaacov@mscc.huji.ac.il. Ya'acov Ritov was partly supported by an ISF grant.}
\thanks{Manuscript received February 7, 2012 ; revised January 8, 2013.
}}

\markboth{IEEE Transactions in Signal Processing}%
{Sepulcre \MakeLowercase{\textit{et al.}}: Sparse regression algorithm for activity estimation in $\gamma $ spectrometry}


\maketitle

\begin{abstract}
We consider the counting rate estimation of an unknown radioactive source, which emits photons at times modeled by an homogeneous Poisson process. A spectrometer converts the energy of incoming photons into electrical pulses, whose number provides a rough estimate of the intensity of the Poisson process. When the activity of the source is high, a physical phenomenon known as pileup effect distorts direct measurements, resulting in a significant bias to the standard estimators of the source activities used so far in the field. We show in this paper that the problem of counting rate estimation can be interpreted as a sparse regression problem. We suggest a post-processed, non-negative, version of the Least Absolute Shrinkage and Selection Operator (LASSO) to estimate the photon arrival times. The main difficulty in this problem is that no theoretical conditions can guarantee consistency in sparsity of LASSO, because the dictionary is not ideal and the signal is sampled. We therefore derive theoretical conditions and bounds which illustrate that the proposed method can none the less provide a good, close to the best attainable, estimate of the counting rate activity. The good performances of the proposed approach are studied on simulations and real datasets.
\end{abstract}


%
\IEEEpeerreviewmaketitle

\section{Introduction}
\label{sec:introduction}

\begin{figure}[ht]
\centering
\includegraphics[width=0.8\linewidth]{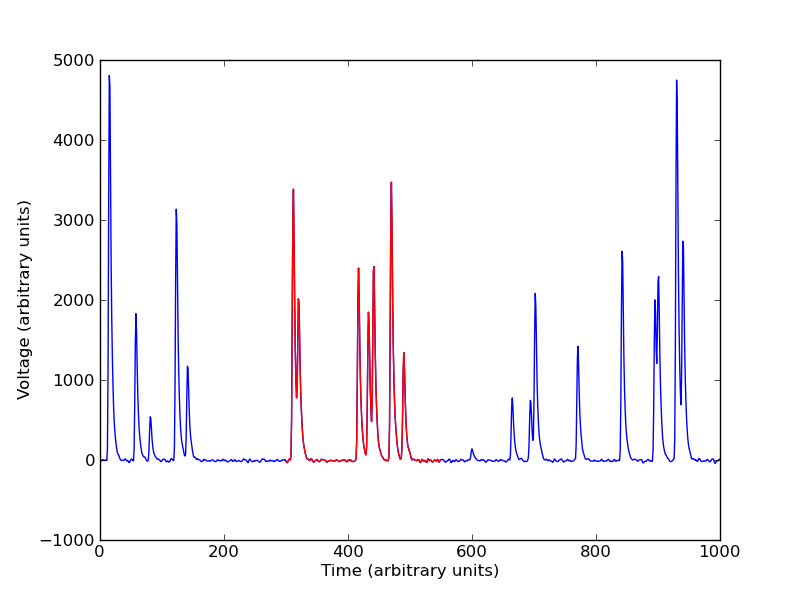}
\caption{Example of a spectrometric signal. The red part is an example of piled up electrical pulses.}
\label{fig:example_pileup}
\end{figure}

\ifx\notIEEE\undefined
\IEEEPARstart{R}{ate}
\else
Rate
\fi
 estimation of a point process is an important problem in nuclear spectroscopy. An unknown radioactive source emits photons at random times, which are modeled by an homogeneous Poisson process. Each photon which interacts with a semiconductor detector produces electron-hole pairs, whose migration generates an electrical pulse of finite duration. We can therefore estimate the activity of the source by counting the number of activity periods of the detector. We refer the reader to~\cite{knoll:1989} and~\cite{leo:1994} for further insights on the physical aspects in this framework. However, when the source is highly radioactive, the durations of the electrical pulses may be longer than their interarrival times, thus the pulses can overlap. In gamma spectrometry, this phenomenon is referred to as pileup. Such a distortion induces an underestimation of the activity, which become more severe as the counting rate increases. This issue is illustrated in~Figure~\ref{fig:example_pileup}.

In its mathematical form, the current intensity as a function of time can be modeled as a general shot-noise process
\begin{equation}
\label{eq:signal-process}
y(t) \eqdef \sum_{k\geq1}E_k \Phi_k(t-T_{k}) \eqsp,
\end{equation}
where $\{E_k,\, k \geq 1\}$ and $\{\Phi_k(s),\, k \geq 1\}$ are respectively the energy and the shape of the electrical pulse associated to the $k$-th photon,
and $y(t)$ defines the continuous time recorded signal. The pulse shapes $\{ \Phi_k(s),\, {k \geq 1} \}$ are assumed to belong to a parametric family of functions $\Gamma_\Theta, \ \Theta \subset \rset^{n}$. The restriction of the signal to a maximal segment where it is strictly positive is referred to as a \textit{busy period}, and where it is 0 as \textit{idle period}. In practice, we observe of sampled version of \eqref{eq:signal-process} with additional noise, and wish to estimate from this recorded digital signal the counting rate activity.

The problem of activity estimation has been extensively studied in the field of nuclear instrumentation since the 1960's (see \cite{bristow:1990} or \cite{michotte:2009} for a detailed review of these early contributions; classical pileup rejection techniques are detailed in \cite{spectro_ansi:1999}). Early papers on pileup correction focus specifically on activity correction methods, such as the VPG (Virtual Pulse Generator) method described in \cite{westphal:1982,westphal:1985}. Moreover, it must be stressed that these techniques are strongly related to the instrumentation used for the experiments. Recent offline methods are based on direct inversion techniques~\cite{trigano:moulines:ieeesp:2006} or computationally intensive methods~\cite{belli:esposito:2008}, and are usually not fitted for very high counting rates. It is of interest to consider fast, event-by-event pile-up correctors for real-time applications, as proposed in~\cite{vencelj:bucar:2009} for calorimetry and in~\cite{vencelj:novak:2009} for scintillators. One of the main advantages of the methods developed in~\cite{trigano:moulines:ieeesp:2006} is that they do not rely on any shape information of the time signal, but rather on the alternance of the idle and busy periods of the detector. However, when the activity of the radioactive source is too high, we observe very few transitions from busy to idle periods, thus making this information statistically irrelevant.

In the latter case, it is therefore necessary to introduce additional assumptions on the pulse shapes (e.g. to specify $\Gamma_\Theta$), and to estimate both the signal sample path on a relevant basis. This can be formally viewed as a regression problem. However, due to the nature of the physical phenomenon, and since Poisson processes usually represent occurrences of rare events, the regressor chosen to estimate the signal must be sparse as well. Since the seminal papers~\cite{tibshirani:1996} and \cite{chen:donoho:saunders:1998}, representation of sparse signals has received a considerable attention, and significant advances have been made both from the theoretical and applied point of view. Several recent contributions~\cite{efron:hastie:johnstone:2004} suggest efficent algorithms yielding estimators with good statistical properties, thus making sparse regression estimators a possible option for real-time processing. In this paper, we chose to use a modification of LASSO with a positivity constraint~\cite{efron:hastie:johnstone:2004}. Indeed, LASSO provides a sparse solution close to the real signal for the $\ell_2$-norm. However, since we are not interested in the reconstruction of the signal for activity estimation, but rather in the Poissonian arrival times, it is of interest to investigate the consistency in selection of the sparsity pattern. Numerous recent works have been devoted to this general question about LASSO, the first ones
being~\cite{zhao:yu:2006} and \cite{meinshausen:buhlmann:2006}. Both papers introduced independently the so-called \textit{irrepresentability} condition as a necessary condition for selection consistency. More recently,~\cite{wainwright:2009} developed the conditions under which the irrepresentability condition is also a sufficient one. We also refer to~\cite{zhang:huang:2008},~\cite{meinshausen:yu:2009},~\cite{vandegeer:buhlmann:2009} and for recent results on consistency in the $\ell_2$-sense for the signal estimation; note however that the estimation of the activity of the source is related to the selection consistency issue, whereas the consistency in the $\ell_2$ sense should be used for energy spectrum reconstruction.
%
%
The problem we address in this paper shares also similarities with the reconstruction of sampled signals with finite rate of innovation~\cite{vetterli_sampling_2002}. In the latter, the authors present a method based on the use of the annihilator filter used in error-correction coding, which allows to reconstruct perfectly a Poisson driven signal made of splines of piecewise polynomials, even when it is not bandwidth limited. This leads to a purely algebraic reconstruction of the signal when it can be decomposed on a known functional base. However, the cornerstone for algebraic reconstruction is the full knowledge of this base, which is not the case in our framework.
The paper is organized as follows. Section~\ref{sec:model_assumptions} presents the model and the derivation of the estimator of the counting rate. This estimation can be roughly seen as a post-processed version of the non-negative LASSO. Though \eqref{eq:signal-process} is rather close to a standard linear regression one, the presented problem is difficult to address, since the discrete signal stemming from $y(t)$ is not generated from a specific, known dictionary. Moreover, it is impossible to infer the exact number of Poissonian arrivals between two sampling points. Both considerations imply that theoretical conditions (e.g. derived in~\cite{wainwright:2009}) which ensure consistency in sparsity are not met in this case.. We therefore present in Section~\ref{sec:theoretical_results} theoretical results showing that the activity of the source can be recovered almost as well as the best estimator we could build from a full knowledge of the Poisson process and discrete observations with a high probability. Finally, section~\ref{sec:applications} illustrates on some applications the effectiveness of the proposed approach, both on simulations and real data, with comments. Details of the calculations and proofs of the presented results are detailed in the appendix. 
\section{Sparse regression based method for activity estimation}
\label{sec:model_assumptions}

\subsection{Model and assumptions}

We observe a signal uniformly sampled on some subdivision $\mathcal{T} \eqdef  \left \{ 0 = t_0 , t_1, t_2 , \ldots ,  t_{N-1}\right \}$ with sampling period $\Delta t$,  
 stemming from a discrete version of \eqref{eq:signal-process}:
\begin{equation}
y_i = \sum_{n=1}^M E_n  \Phi_n (t_i - T_n) +  \varepsilon_i,\ \ \ 0 \leq i \leq N-1\, , \label{eq:discrete_model}
\end{equation}
where $\left \{ T_n \, , \, 1 \le n \le M \right \}$ is the sample path of an homogeneous Poisson process with constant unknown intensity $\lambda$, $\{E_{n} , 1\le \ n \le M \}$ is a sequence of independent and identically distributed (iid) random variables representing the photons energies, with unknown probability density function $f$, $\{\Phi_{n} ,\  1\le n \le M \}$ is a sequence of functions to be defined later which characterize the electric pulse shapes generated by the photons, and $\{\varepsilon_i, \, 0\le  i \leq N-1 \}$ is a sequence of iid Gaussian random variables with zero mean and variance $\sigma^2$ representing the additional noise of the input signal. Alternatively, when defining the matrix $\boldsymbol{\Phi} \eqdef \left [ \Phi_{n} (t_{i} - T_{n}) \right ]_{0 \leq i \leq N-1 , 1 \leq n \leq M}$ and the vectors $\mathbf{y} \eqdef \left [ y_0, y_1 , \ldots ,  y_{N-1}  \right ]^{T}$, $\mathbf{E} \eqdef [E_{1} , \ldots , E_{M}]^{T}$ and ${\boldsymbol{\varepsilon}} \eqdef [\varepsilon_0, \ldots, \varepsilon_{N-1} ]^T$,\eqref{eq:discrete_model} can be rewritten in a matricial form:
\begin{equation}
\mathbf{y}= \boldsymbol{\Phi} \mathbf{E} + \boldsymbol{\varepsilon} \ .
\label{eq:linear_model_matrix_random}
\end{equation}
All along the paper, it is assumed for convenience that $N$ is an even number. The problem to address is the estimation of $\lambda$ given $\mathbf{y}$. However, no $T_n$ belongs to $\mathcal{T}$ with probability 1. 
We thus introduce the following integer subset related to the closest sample times from the Poisson arrivals $T_n$: 
\begin{equation}
\label{eq:poissonroundings}
P_{0} \eqdef \{  \lfloor T_{n} / \Delta t \rfloor; \, \, n=1,\ldots,M    \},
\end{equation}
where $\lfloor x \rfloor$ denotes the closest integer to $x$. Note that provided $\lambda \Delta t  \ll 1$,  $P_{0}$  is a sparse subset of $\{0,1,\ldots,N-1\}$. We further on denote by $\overline{\mathbf{y}}$ the noise-free part of the signal \eqref{eq:linear_model_matrix_random}, that is  $\overline{\mathbf{y}} \eqdef \boldsymbol{\Phi}\mathbf{E}$. All along the paper, it is assumed that the random variables $\{ E_{n}, \, 1 \leq n \leq M \}$ are bounded by positive and known constants $E_{\min},E_{\max}$:
\begin{equation}
\label{eq:positive_bounds_levels_energies} 0 < E_{\min} \leq E_{n} \leq E_{\max}, \ n=1,\ldots,M \eqsp .
\end{equation} 
In practice, neither $\mathbf{E}$ nor $\boldsymbol{\Phi}$ are known, so the problem cannot be seen as a standard regression problem. Nevertheless, a single electrical pulse $\Phi_{n}$ has a specific shape, characterized in most detectors by a rapid growth created by the charge collection followed by an exponential decay as the charges migrate to the detector electrodes. Thus, a natural idea is to rely on some user predefined dictionary of truncated gamma shapes in order to obtain a modelization of \eqref{eq:linear_model_matrix_random} we can work with. Since a gamma shape is parametrized by two scale and shape parameters, we define a set of $p$ pairs of such parameters by $\boldsymbol{\theta} \eqdef  \{ (\theta_{1}^{(s)}, \theta_{2}^{(s)}) ; \, \, s=1,2,\cdots,p \}$. For all $s=1,\ldots,p$, we define the following pulse shape:
$$
\Gamma_{s}  (t) \eqdef c_{s} \, t^{\, \theta_{1}^{(s)}} \exp( - \theta_{2}^{(s)} \, t ) \, \1 (0<t \leq\tau \Delta t ),
$$
where $\tau$ is a positive constant integer defining the common support of the pulse shapes, and $c_{s}$ is a normalizing constant chosen so that $\frac1N \sum_{i=0}^{N-1} \Gamma_{s}  (t_i)^2 = 1$. Accordingly, we define the following $N \times p$ matrix $\mathbf{A}_k$ whose columns are sampled versions of the previously defined pulse shapes, translated by $t_k$:
\begin{equation}
\mathbf{A}_{k} \eqdef \left [ \Gamma_s (t_i- t_k ) \right ]_{0\leq i \leq N-1,1\leq s \leq p}. \label{eq:timeblock} 
\end{equation}
We further on refer to $\mathbf{A}_{k}$ in \eqref{eq:timeblock} as the \textit{time block} associated to the $k$-th point. We now define a global dictionary $\textbf{A}$ by concatenating these time blocks:
\begin{equation}
\textbf{A}=\left[\textbf{A}_{0} \ \  \textbf{A}_1 \ \  \cdots \ \  \textbf{A}_{N-1} \right] \ .
\label{eq:global_dictionary}
\end{equation}
Note that \eqref{eq:global_dictionary} defines a $N\times Np$ matrix with full rank $N$. Therefore, an equivalent version of  \eqref{eq:linear_model_matrix_random} consists in some linear decomposition of $\overline{\mathbf{y}}$ along the columns of $\textbf{A}$, for some unknown regressor $\boldsymbol\beta$. Recall that with probability $1$,  no $T_n$ belongs to $\mathcal{T}$ neither $\Phi_n$ is a column of $\mathbf{A}$. In that sense, we shall say that our dictionary $\mathbf{A}$ is \emph{incomplete}. 
Therefore, the model investigated throughout the rest of the paper is
\begin{equation}
\label{eq:model_approx}
\mathbf{y}= \mathbf{A} \boldsymbol\beta + \boldsymbol\delta + \boldsymbol\varepsilon ,
\end{equation}
in which $\boldsymbol\delta \eqdef \overline{\mathbf{y}} - \mathbf{A} \boldsymbol\beta $ denotes the discrepency between the decomposition on $\boldsymbol\Phi$ and $\mathbf{A}$, and which $\textbf{A}  \boldsymbol\beta $ belongs to the closed positive span $\mathcal{C}$ defined as
\begin{multline}
\label{eq:pos_cone}
\mathcal{C} \eqdef \Bigl \{ \sum_{i\in P_{0}} \mathbf{A}_{i} x_{i}\ ; \ x_{i} \in \mathbb{R}_{+}^{p}  \text{ such that } \\ E_{\min} \leq  \frac{1}{\sqrt{N}} \| \mathbf{A}_{i} x_{i} \|_{2} \leq E_{\max} \Bigr \} \ .
\end{multline}
Note that $\mathcal{C}$ parametrizes models supported on timeblocks indexed by $P_{0}$ only. 
The reason for bounding $ \left \| \textbf{A}_{i} \, x_{i} \right \|_{2} $ in \eqref{eq:pos_cone} can be understood in light of~\eqref{eq:positive_bounds_levels_energies}. Since $P_{0}$ contains essentially all the information one could ever retrieve from $T_n$, the set of all the $T_n$ and $\Delta t P_0$ tend to be identical as $\Delta t $ tends to $0$. Since we expect the decomposition $\mathbf{A} \boldsymbol\beta $ to be quite close to the decomposition  $ \boldsymbol\Phi \mathbf{E} $, it is rather natural to focus on the best discrepancy measure $\boldsymbol\delta$ one could ever get when imposing similar constraints on the energies encoded in the model $\mathbf{A} \boldsymbol\beta $. Following \eqref{eq:model_approx}, and since we want to focus on model with small discrepancy, we use in the rest of the paper $\alpha \eqdef \|\boldsymbol\delta\|_2 / \sqrt{N}$ as a standard discrepancy measure.




%
%
%
%
%
%
%
%
\subsection{Additional notations}
\label{subsec:notas}

We introduce here for convenience the notations used in the following sections. Given any finite discrete set $I$, we denote by $|I|$ its cardinality. We denote by $\1_{I}$, the column vector of length $|I|$ whose all coefficients are equal to $1$. If $\textbf{u},\textbf{v}$ are two vectors of identical size, we shall write $\textbf{u} < \textbf{v}$ (respectively $\textbf{u} \leq \textbf{v}$) when all entries of $\mathbf{v}-\mathbf{u}$ are positive (nonnegative). 

 If $n$ is any integer in $\{ 1 , \cdots, N p \}$, thus indexing a column of $\ab$, we shall refer to this column by $A_{n}$; similarly for any regressor $\bbe \in \rset^{N p}$ the $n$th entry is denoted by $\beta_{n}$. 
Given $I$ a subset of $\{0,1,\ldots,N-1\}$, we denote by $\mathbf{A}_I$ the submatrix obtained by concatenation of times blocks whose index belong to $I$, namely 
\begin{equation}
\label{eq:subdicoI}
\mathbf{A}_{I} \eqdef [\cdots, \mathbf{A}_{k} , \cdots ]_{k \in I}\ .
\end{equation}
Given two subsets $I,J$ of $\{0,1,\ldots,N-1\}$, we define the Gram matrix of size $p |I| \times p |J|$ associated to $A_I$ and $A_J$ as
\begin{equation}
\label{eq:def_Grammatrix_between_blocks} 
\mathbf{G}_{I, J} \eqdef \frac{1}{N} \mathbf{A}_{I}^T \mathbf{A}_{J},
\end{equation}
and whenever $I=J$ and reduces to one singleton, we shall drop the dependency in $I,J$ and write more simply $\mathbf{G}$, since by the very construction of $\mathbf{A}$, the Gram matrix of one timeblock is independent of the block index.

Given $\boldsymbol\beta$ a vector of size $ pN  $, it will be naturally decomposed along the timeblocks:
$\boldsymbol{\beta} =  [ \boldsymbol\beta_{0}^T ; \cdots ; \boldsymbol\beta_{N-1}^T ]^{T}$, where for all $i$, $ \boldsymbol\beta_{i} \in \rset^{p}$. We define the block pattern of $\boldsymbol{\beta}$ as 
\begin{equation} 
\label{eq:tbpattern} 
J( \boldsymbol{\beta} ) \eqdef \{ i \ ; \ \boldsymbol{\beta}_{i} \neq 0 \} \ .
\end{equation} 
Given some integer $k$ in $\{0,1,\ldots,N-1\}$, we define for all positive real $\alpha$ the $\alpha$-\textit{neighborhood} of $k$ as: 
\begin{equation} \label{eq:neighbi}
V_{\alpha}(k) \eqdef [k - \alpha \, ; \, k + \alpha] \cap \{0,1,\ldots,N-1\},
\end{equation}
and denotes his complement by $\overline{V_\alpha (k)} =\{0,1,\ldots,N-1\} \setminus V_{\alpha}(k)$. Alternatively, \eqref{eq:neighbi}  can be reformulated accordingly to the correlations between blocks, since the discrete correlation between two shapes is a decreasing function of the distance between their time shifts, and is zero whenever this distance is greater than $\tau$. Therefore, given a integer $k$ in $\{0,1,\ldots,N-1\}$, we define some neighbourhood of $k$ accordingly to some specified correlation level $0\leq \rho \leq 1$:
\begin{equation} 
\label{eq:neighb_correl} 
\mathcal{T}_{\rho}(k) \eqdef \{0,1,\ldots,N-1\} \setminus \{ j\, ; \ \max |\mathbf{G}_{\{k\},\{j\}} | < \rho \} .
\end{equation}
In other words, $j \in \overline{\mathcal{T}_{\rho}(k)} $ if and only if every column in the $j$-th timeblock has a correlation lower than $\rho$ with every column in $\textbf{A}_{k}$, expressing the fact that $t_{j}$ is somehow 'distant' from $t_k$. Due to these considerations, we can associate to $\rho \in [0,1]$ a real $\alpha$ such that $V_{\alpha}(k) = \mathcal{T}_{\rho}(k)$. Obviously, when $\rho^{'} \leq \rho$ one has $\mathcal{T}_{\rho}(k)  \subset \mathcal{T}_{\rho'}(k)  $, that is $\alpha \leq \alpha'$. 

We eventually define two quantities which will appear in the theoretical bounds obtained. If $k$ is an integer such that $\tau \leq k \leq N-\tau-1$, we shall define
\begin{equation} \label{eq:gsumentriesgram} 
\mathcal{G} \eqdef \frac1N \sum_{l \in V_{\tau}(k)} \max_{i,j} \mathbf{A}_k^T\mathbf{A}_l (i,j)
\end{equation}
as the sum of all maximal correlations per block with respect to the $k$-th timeblock (note that $\mathcal{G} $ is independent of $k$), and
\begin{equation*}
\mathfrak{t}(x) \eqdef \frac{1}{x\sqrt{2\pi}}e^{-x^2/2}
\end{equation*}

\subsection{Overview of the estimation procedure}
\label{subsection:procedure_settings}
Recall that our objective is to estimate $\lambda$ given $\textbf{y}$. It is well known that if $\{T_n,\ 1\leq n \leq M\}$ are the points of an homogeneous Poisson process, the inter-arrival times are iid random variables with common exponential distribution with parameter $\lambda$. Therefore, $\lambda$ can be consistently estimated by
\begin{equation}
\displaystyle \lambda_{c} \eqdef \frac{M}{ T_M } \eqsp .
\label{eq:activity_ideal_estimate}
\end{equation}
However, $\mathbf{y}$ is a discrete-time signal, therefore \eqref{eq:activity_ideal_estimate} cannot be attained since we are restricted to use only times in $\mathcal{T}$. Therefore, the best estimate of $\lambda$ attainable in practice is defined as
\begin{equation}
\lambda_\text{opt} \eqdef \frac{|P_0|}{\Delta t \max P_0} \ . \label{eq:activity_optimal_estimate}
\end{equation}
%
It is likely that $\lambda_\text{opt} < \lambda_{c}$, since $|P_0| < M$; however, provided $\lambda \Delta t$ is small, $\lambda_c$ and $\lambda_{\text{opt}}$ should remain close. The main idea of this paper is therefore to plug in \eqref{eq:activity_optimal_estimate} estimates of $M$ and $T_M$, as now explained. If the signal is modelled by \eqref{eq:model_approx}, the set $J( \boldsymbol\beta )$ still contains much fewer elements than $N$. Thus, we would like to recover first $J( \boldsymbol{\beta} ) $, and make use of a non-negative LASSO estimator (NNLASSO)~\cite{tibshirani:1996,efron:hastie:johnstone:2004}:
\begin{align}
\widehat{\boldsymbol\beta}(r) &=  \underset{ \boldsymbol\beta \in \mathbb{R}^{N  p}  }{\argmin}\ \Bigl\{ \frac{1}{2 N } \left \Vert \mathbf{y} - \sum_{m=0}^{N-1} \mathbf{A}_{m} \,   \boldsymbol\beta_{m} \right \|_2^2  + r \, \sum_{m=0}^{N-1} |\boldsymbol\beta_{m} |_{\ell_1} \Bigr\} \label{eq:estimator_general_def} \\
& \text{ such that } \boldsymbol{\beta} \geq 0 , \nonumber 
\end{align}
where the tuning parameter $r$ quantifies the tradeoff between sparsity and estimation precision. NNLASSO provides a sparse estimator $[ \widehat{\boldsymbol\beta}_{0}^T(r) , \cdots , \widehat{\boldsymbol\beta}_{N-1}^T (r)]^T $ such that the linear model $\widehat{\mathbf{y}} = \mathbf{A} \widehat{ \boldsymbol\beta }(r)$ approximates accurately the signal $\mathbf{y}$.  In practice, \eqref{eq:estimator_general_def} can be efficiently computed by a modification of the LARS algorithm~\cite{efron:hastie:johnstone:2004}.
%
Note that the group-LASSO~\cite{yuan:lin:2006} also exploits the time blocks decomposition of $\boldsymbol\beta$ and provide a block-sparse regressor. However, in this paper, we cannot assume the groups to be fully known, due to the incompleteness of $\ab$.

Assuming that the solution \eqref{eq:estimator_general_def} is known, the estimation of $\lambda$ should be carefully done. 
 It is tempting to estimate the arrival times with the set $J(\widehat{\boldsymbol\beta }(r))$ and the total number of occurrences by $ |J(\widehat{\boldsymbol\beta }(r))|$, then plug this data into \eqref{eq:activity_optimal_estimate}. However $J(\widehat{\boldsymbol\beta }(r))$ may contain consecutive active time blocks which do not all correspond to real arrival times. This is not surprising: since $\mathbf{A}$ is incomplete, $J(\boldsymbol\beta ) $ may itself be distinct from $P_0 $ . 
In this paper we suggest an additional thresholding step to the estimation of $\lambda$ to circumvent this issue, that is
\begin{enumerate}
\item solve~(\ref{eq:estimator_general_def}) to obtain $\widehat{\boldsymbol\beta}(r)$.
\item set all the $\widehat{\boldsymbol\beta}_m (r)$ such that $\|\widehat{\boldsymbol\beta}_m (r)\|_1 < \eta$ to zero, where $\eta$ is a user defined threshold to be precised later;
\item estimate the arrival times recursively $\widehat{T}_n \eqdef \underset{k=0,\ldots,N-1}{\min} \{k\Delta t > \widehat{T}_{n-1} \ ; \ \widehat{\boldsymbol\beta}_{k-1}(r) = 0,  \widehat{\boldsymbol\beta}_{k}(r) \neq 0 \}$, and $\widehat{M} \eqdef | \{k\ ;\ \widehat{\boldsymbol\beta}_{k-1} (r) = 0, \widehat{\boldsymbol\beta}_{k}(r) \neq 0\} |$.
\item Estimate the activity as
\begin{equation} \label{eq:Lasso_activity_estimate}
 \widehat{\lambda}(r,\eta)  \eqdef \frac{\widehat{M}}{\widehat{T}_{\widehat{M}}}  
 \end{equation}
\end{enumerate}
The pseudocode of the latter algorithm is summarized in Algorithm~\ref{alg:count_rate_LASSO}.
\begin{algorithm}
\label{alg:count_rate_LASSO}
\DontPrintSemicolon
\SetKwInOut{Input}{Input}
\SetKwInOut{Output}{Output}
\Input{Input signal $\mathbf{y}$, Dictionary $\mathbf{A}$, Sparsity parameter $r$, Threshold $\eta$.}
\Output{Intensity estimate $\widehat{\lambda}(r,\eta)$.}
\Begin{
{\bf solve} \begin{multline*}
\widehat{\boldsymbol\beta}(r) =  \\ \underset{ \boldsymbol\beta \in \mathbb{R}^{N  p}, \boldsymbol\beta \geq 0  }{\argmin}\ \Bigl\{ \frac{1}{2 N } \Bigl\| \mathbf{y} - \sum_{m=0}^{N-1} \mathbf{A}_{m} \,   \boldsymbol\beta_{m} \Bigr\|_2^2  + r \, \sum_{m=0}^{N-1} |\boldsymbol\beta_{m} |_{\ell_1} \Bigr\}
\end{multline*}\;
{\bf set} $M=0$, $T=0$\;
\For{$1 \le m \le N-1$}{\If{$\|\widehat{\boldsymbol\beta}_m (r)\|_1 >  \eta$ {\bf and} $\|\widehat{\boldsymbol\beta}_{m-1} (r)\|_1 < \eta$}{
{\bf set} $T \leftarrow m\Delta t$\;
{\bf set} $M \leftarrow M+1$\;
}
}
{\bf compute}  $\widehat{\lambda}(r,\eta) = \dfrac{T}{M}$\;
}
\caption{Intensity estimation by post-processed NNLASSO}
\end{algorithm}

We refer to steps 2 and 3 as "post processing" steps. Both steps can be heuristically understood as follows: step 2 in introduced since time blocks containing negligible weights are probably selected to improve slightly the estimation, but are not related to pulses start; indeed in realistic situations all the pulses considered, including the real ones, start with similar sharp slopes, but decrease differently, which makes these "negligible" time blocks appear behind the pulse start. In step 3 we merge consecutive selected time blocks due to high correlations between blocks and incompleteness of the dictionary, as mentioned above. Clearly a good estimation of $\lambda$ is conditioned by a careful choice of both sparsity and thresholding parameters $r$ and $\eta$. A reasonable a practical choice is to set them accordingly to the noise variance, as seen in the applications section. Note also that the cornerstone of Step 2 is that $\lambda$ is small enough with respect to the signal sampling frequency, so that consecutive active blocks would unlikely correspond to two distinct events. Even if this thresholding fails in case of extremely high counting rates, as seen in the application sections, we emphasize that it covers most spectrometric applications, making it very relevant in practice.

%
%
Note also that alternative methods to \eqref{eq:estimator_general_def}, which are based on iterative and reweighting procedures, exist~\cite{candes_enhancing_2008}. These methods seem appealing for higher activities, since they are known to provide sparser solutions than NNLASSO. Similarly, sparse Bayesian learning techniques~\cite{wipf_iterative_2010,wipf_nips_2011} are known to provide sparser results in the case of very correlated dictionaries. Nevertheless, in practice, the fact that the spectrometric signal $\mathbf{y}$ does not stem directly from $\textbf{A}$ cripples their performances, and the post-processing introduced in the latter remains necessary even in this case, as seen later in the applications section.
%
%
%



\section{Theoretical results}
\label{sec:theoretical_results}

In order to guarantee some consistency in estimation as well as in selection, previous works imposed conditions on the dictionary $\textbf{A}$: for instance low correlations between columns~\cite{donoho:huo:2001,bickel:ritov:tsybakov:2008,lounici:2008,ben-haim:eldar:elad:2010} or positivity of minors of specific sizes~\cite{meinshausen:yu:2009,bickel:ritov:tsybakov:2008,zhang:huang:2008}. The estimation procedure described in this paper is close to~\cite{meinshausen:yu:2009}, which suggest improvements of LASSO by hard-thresholding coefficients, so that only representative variables are selected.
In~\cite{zhao:yu:2006,wainwright:2009}, the so called {\it{irrepresentability}} condition is introduced, and is proved to be necessary if we wish selection consistency with high confidence. 
In our case remember we wish to recover $P_{0}$, thus the next subsection details an irrepresentability condition on $\underset{ \lVert z \rVert_{\infty} \leq 1 }{\max } \lVert \mathbf{G}_{\overline{P_{0}}, P_{0}} \, \mathbf{G}_{P_{0} , P_{0}}^{-1} z \rVert_{\infty} $ so that NNLASSO could theoretically select variables belonging only to $P_{0}$. Nevertheless, this insight is not relevant practically. Therefore, further theorems~\ref{th:bounding_intensity_lambdaP} and~\ref{cor:relation_lambdaopt_lambdapost} do not use this assumption, and rather compare the actual timeblock pattern and the one obtained with NNLASSO in terms of intersecting neighborhoods.

\subsection{Exact timeblock recovery and bounds for sparsity pattern approximation}    
 \label{subsec:exacttbrecov}

%
For any value of the parameter $r >0$, recall that we defined $\widehat{\boldsymbol\beta} (r) $ as the NNLASSO minimizer \eqref{eq:estimator_general_def}. Next proposition is closely adapted from~\cite{wainwright:2009}, and provides, under very specific conditions on the dictionary $\mathbf{A}$, some range of values of $r$ such that the NNLASSO minimizer selects only time blocks from $P_0$.
\begin{prop}
\label{prop:first_question}
Assume that for all vectors $\textbf{z}$ of length $| P_0 |$ such that $ \textbf{z} \leq \1_{\po} $ the following assumption holds:
 \begin{equation}
 \label{eq:block_irrep_cond}
\mathbf{G}_{\overline{P_{0}}, P_{0}} \, \mathbf{G}_{P_{0} , P_{0}}^{-1} \textbf{z}   < (1- \eta_{0} ) \1_{\overline{\po}} 
 \end{equation}
 for some $0<\eta_{0} <1 $.  If the parameter $r$ is chosen such that
\begin{equation}\label{eq:choice_parameter}
    r > \max \left \{ \frac{2 \alpha}{ \eta_0} \, ; \, \frac{ 2 \, \sqrt{2} \, \sigma}{\eta_0} \, \sqrt{\frac{\log (N - | P_0 | )p}{N}} \right \} \, ,
\end{equation}
then $\widehat{\boldsymbol\beta} (r) $ is supported by $P_0$ with
 probability tending to $1$ as $N$ tends to infinity.
\end{prop}
\begin{pv} 
See Appendix~\ref{sec:proof_q1}.
\end{pv}
%
%

Though similar to standard conditions which ensure the consistency of LASSO appearing e.g. in~\cite{wainwright:2009}, Proposition~\ref{prop:first_question} is of little practical use, since the dictionary~$\mathbf{A}$ does not satisfy the latter conditions. In light of~\eqref{eq:choice_parameter}, we can also remark that a good choice of the sparsity parameter depends on many terms ($\alpha,\ P_{0}$) unknown in practice. This illustrates the need of further results, since the question arose in this paper is whether and in which measure standard sparse methods could provide sufficiently good results even when theoretical conditions are not met. 
%
The two main theorems are based on the following proposition\ref{prop:bounds_sparsitylevel_LASSOregressor}, whose aim is to compare
the true sparsity pattern with the NNLASSO one, for convenient choice of $r$, and adequate block thresholding. The short technical lemma
 \eqref{lem:bounding_l1normblocks_model}, which is a direct consequence of the definition of $\mathcal{C}$, see \eqref{eq:pos_cone}, will be used in the proof.
 
\begin{prop}\label{prop:bounds_sparsitylevel_LASSOregressor}
Define the following threshold value
\begin{equation}
\label{eq:thresholdvalue}
\eta \eqdef \frac{ E_{\min}^{2} \, \min_{i,j} \bg(i,j)^{1/2}}{4 (2 \tau +1) E_{\max}},
\end{equation}
and assume that $r$ satisfies
\begin{equation}
\label{eq:r_choice}
r + \alpha < \frac{ E_{\min}^{2} \, \min_{i,j} \bg(i,j)^{1/2} }{2 \, E_{\max}} \ .
\end{equation}
Then, there exists $0\le \rho \le 1$ (dependent on $r$ and $\eta$, but independent of $\bbe$) such that, for all $k$ in $P_{0}$, there is an integer $m$ in $\mathcal{T}_{\rho}(k)$ so that $ \| \wbbe_{m} \|_{1} \geq \eta$ with probability greater than
\begin{multline}
\label{eq:nonzerothresholdblocklasso}
1 - p \mathfrak{t} \left ( \sqrt{N} \, \frac{E_{\min}^{2} \, \min_{i,j} \bg(i,j)^{1/2}}{4 \, E_{\max} \, \sigma} \right ) \\ - p (2 \tau +1 ) \mathfrak{t} \left ( \sqrt{N}  \, \frac{r - \alpha}{\sigma} \right ) \ .
\end{multline}
Conversely, there exists $0\le \mu \le 1$ (dependent on $\eta$, but independent from $\bbe$ and $r$) such that, for any $k$ satisfying $ \| \wbbe_{k} \|_{1} \geq \eta $ we have: 
\begin{equation}
\label{eq:nonzeroactualblock} 
\Pr ( P_{0} \cap \mathcal{T}_{\mu}(k) \neq \emptyset ) > 1 -  \| \wbbe_{k} \|_{0} \, \mathfrak{t} \left ( \sqrt{N} \, \frac{r - \alpha}{\sigma} \right ) .
\end{equation}
\end{prop}
\begin{pv}
See Appendix~\ref{proof:bounds_sparsitylevel_LASSOregressor}.
\end{pv}  

Proposition~\ref{prop:bounds_sparsitylevel_LASSOregressor} is of practical interest. Roughly speaking, it states that provided the threshold parameter of the post-processing steps and the sparsity parameter are set accordingly to~\eqref{eq:thresholdvalue} and~\eqref{eq:r_choice}, then any element of the close to optimal sparsity pattern $P_0$ has in his close vincinity an integer $k$ so that $\|\widehat{\bbe}_k\|_1$ is selected in step 2 of our algorithm, and conversely. Therefore, the latter result closely relates $J(\widehat{\bbe})$ after post-processing to $P_0$, thus connecting $\widehat\lambda(r,\eta)$ to $\lambda_{opt}$. It is clear that in practice, the value of $\alpha$ is unknown. However, the results still stands for smaller values of $\eta$, such as the one selected in the application section.



\subsection{Confidence bounds for counting rate estimators}

The two next theorems are strongly based on Proposition~\ref{prop:bounds_sparsitylevel_LASSOregressor}. They provide computable bounds of confidence intervals for $\widehat{\lambda}(r,\eta)- \lambda_{opt}$, where $\widehat\lambda(r,\eta)$ and $\lambda_{opt}$ were respectively defined in~\eqref{eq:Lasso_activity_estimate} and~\eqref{eq:activity_optimal_estimate}. Recall that only the blocks of $\wbbe (r)$ selected by the criterion $\| \wbbe_{k} (r)  \|_{1} > \eta $ are used to estimate the arrival times. 
In the next theorems, we define for brevity $\wbbe(r,\eta)$ as the obtained vector after post-processing, that is
$$
\wbbe(r,\eta) \eqdef [\wbbe_k^T(r) 1(\| \wbbe_{k} (r)  \|_{1} > \eta)]_{0\le k \le N-1}^T
$$
and for any subset $X$ of $\{0,1,\ldots,N-1\}$ defined as a union of discrete intervals, we denote by $I (X)$ the number of these intervals.
%
\begin{theo}
\label{th:bounding_intensity_lambdaP}
Under the same assumptions and settings as in Proposition~\ref{prop:bounds_sparsitylevel_LASSOregressor}, and define $a_{\rho}, a_{\mu}$ as the integers satisfying $V_{a_{\rho}}(k) = \mathcal{T}_{\rho}(k)$, $V_{a_{\mu}}(k) = \mathcal{T}_{\mu}(k)$, for any integer $k$ in $[\tau , N-1-\tau]$ (to avoid interval truncature). We get that
\begin{multline}\label{eq:lambdaoptupbound} 
\widehat\lambda(r,\eta) - \lambda_{opt} \leq \widehat\lambda(r,\eta) \\ \times \left [ 1 - \frac{\widehat{T}_{\widehat{M}} / \Delta t }{a_{\rho} + \max J(\wbbe(r,\eta))} 
 \frac{I \Bigl ( \underset{1 \leq j \leq \widehat{M}}{\bigcup} V_{a_{\mu}} (\widehat{T}_{j}  ) \Bigr )}{\widehat{M}} \right ]
 \end{multline}
 with probability greater than
\begin{multline*}
1 -  p \mathfrak{t} \left ( \sqrt{N} \, \frac{E_{\min}^{2} \, \min_{i,j} \bg(i,j)^{1/2}}{4 \, E_{\max} \, \sigma} \right )\\  - p (2 \tau +1 ) \mathfrak{t} \left ( \sqrt{N}  \, \frac{r - \alpha}{\sigma} \right ) -  \| \wbbe(r)  \|_{0} \, \mathfrak{t} \left ( \sqrt{N} \, \frac{r - \alpha}{\sigma} \right ) 
\end{multline*}

 \end{theo}
\begin{pv}
See appendix \ref{app:bounding_intensity_lambdaP}.
\end{pv} 
 
Roughly speaking, Theorem~\ref{th:bounding_intensity_lambdaP} gives a lower bound to $\widehat\lambda(r,\eta)-\lambda_{opt}$ in terms of the NNLASSO sparsity block pattern, making it numerically computable. It is straightforward to see that the term between brackets in the lower bound of~\eqref{eq:lambdaoptupbound} is positive. The integers $\widehat{T}_{\widehat{M}} / \Delta t$, $\max J(\wbbe(r,\eta))$ are the extremities of the last interval in $ J(\wbbe(r,\eta))$, thus the shorter this interval is the less underestimated $\lambda_{opt}$ is. Moreover, the term $\widehat{M}^{-1}I \Bigl ( \bigcup_{1 \leq j \leq \widehat{M}} V_{a_{\mu}} (\widehat{T}_{j} ) \Bigr )$ is equal to one as soon as $\widehat{T}_{i} - \widehat{T}_{i-1} > a_{\mu} $ for all $1 < i \leq \widehat{M}$, that is when the estimated times are sufficiently spaced. Note that this justifies our choice to define the $\widehat{T}_{i}$'s as the beginnings of the open components, since times belonging to such an open component are not likely to have separated $a_{\mu}-$neighbourhoods. Thus, in the most favorable case, $\widehat{T}_{\widehat{M}} = \Delta t \max J(\wbbe(r,\eta))$ and the number of components is equal to $\widehat{M}$, in other words
\begin{equation*}
\widehat\lambda(r,\eta) - \lambda_{opt} \leq \lambda(r,\eta) \left ( 1 + \frac{\max J(\wbbe(r,\eta))}{a_{\rho}} \right )^{-1},
\end{equation*}
thus showing that $\widehat\lambda(r,\eta)$ is close to $\lambda_{opt}$ with a high probability. The asymptotic study as well as theoretical insights on the lengths of intervals inside the NNLASSO timeblock pattern, in terms of $r$ and other quantities involved in the problem, are beyond the scope of the present paper. The next theorem provides a computable lower bound for $\widehat\lambda(r,\eta)-\lambda_{opt}$.


\begin{theo}\label{cor:relation_lambdaopt_lambdapost}
Under the same assumptions and conventions used in Theorem~\ref{th:bounding_intensity_lambdaP}, suppose moreover that 
\begin{equation}
\label{eq:proba_distancebetweentimes} 
(\lambda \Delta t)^{2} N \, a_{\rho} < 1 .
\end{equation}
Then
\begin{multline}
\label{eq:upperbound_lambdaopt}
 \widehat\lambda (r,\eta) - \lambda_{opt} \geq \widehat\lambda (r , \eta) \\ \times \left [ 1 - \frac{|J(\wbbe(r,\eta))|}{\widehat{M}} \cdot \frac{\widehat{T}_{\widehat{M}} / \Delta t}{\max J(\wbbe(r,\eta)) - a_{\mu} } \right ] 
 \end{multline}
with probability greater than 
\begin{multline*}
1   -   (\lambda \Delta t)^{2} N \, a_{\rho}  - |P_{0}| \Biggl [ p \mathfrak{t} \Bigl( \sqrt{N} \, \frac{E_{\min}^{2} \, \min_{i,j} \bg(i,j)^{1/2}}{4 \, E_{\max} \, \sigma} \Bigr ) \\ - ( p (2 \tau +1 ) +  \| \wbbe(r)   \|_{0} )\mathfrak{t} \left ( \sqrt{N} \, \frac{r - \alpha}{\sigma} \right ) \Biggr ] .
\end{multline*}
 \end{theo}
\begin{pv}
See appendix \ref{app:bounding_intensity_lambdaP}.
\end{pv}

In the latter, the quotient $\frac{|J(\wbbe(r,\eta))|}{\widehat{M}} $ in \eqref{eq:upperbound_lambdaopt} can be interpreted as some average number of active consecutive blocks after thresholding. Note that the probability given here is lower than the one in Theorem~\ref{th:bounding_intensity_lambdaP}.

\section{Applications}
\label{sec:applications}

We present in this section results on realistic simulations, which emphasize the effectiveness of the proposed approach when compared to a standard method (comparison to a fixed threshold and estimation of $\lambda$ by means of the idle times of the detector, see~\cite{trigano:moulines:ieeesp:2006}). The proposed algorithm for counting rate estimation is then studied on a real dataset.

\subsection{Results on simulations}

\subsubsection{Experimental settings}

The performances of the proposed approach are investigated for $ 50 $ points of an homogeneous Poisson process whose intensity $ \lambda $ varies from $ 0.1 $ to $ 0.4 $, which corresponds to physical activities from $ 1.10^6 $ and up to $ 4.10^6 $ photons per second when the signal is sampled to $10$ MHz. Those numbers are related to high or very high radioactive activities, as mentioned for example in~\cite{spectro_ansi:1999}. The energies $ \{E_n, n \ge 0 \} $ are drawn accordingly to a Gaussian density truncated at $0$, with mean $ 50 $ and variance $ 5 $. We present both results in the case of a good Signal Noise Ratio ($ \sigma = 1$), as can be found in Gamma spectrometry applications.

Assuming that we observe $ N $ points of the sample signal, the $ j $-th column of the dictionary $\mathbf{A}$ is build accordingly to~\eqref{eq:global_dictionary}. The grid $\boldsymbol{\theta}$ is taken uniform on $(0,10]^2$, with subdivision step $0.1$. In order to check the robustness of the approach and its practical implementation for real-time instrumentation, the signals are simulated in two different settings:
\begin{itemize}
\item for each point of the Poisson process, a shape is taken randomly from the dictionary $A$; this case will later on be denoted by (I).
\item for each point of the Poisson process, a shifted Gamma is created with randomly chosen parameters $\theta_1,\theta_2$. In our experiments, both parameters are drawn uniformly accordingly to $\theta_1 \sim U([0;10])$ and $\theta_2 \sim U([0;2])$  (case denoted by (II))
\end{itemize}
It is obvious that the standard framework for regression is (I); however, as mentioned earlier, we also want to investigate how the algorithm behaves when the dictionary is not rich enough to cover all the possible shapes, and check the effectiveness of the additional post-processing step introduced in the latter sections. This allows to use the proposed approach on real-world experiments where fast algorithms and small dictionaries for real-time implementations must be used. For one given activity, the estimator is computed $ 10000 $ times by means of the proposed method, and by means of the standard method aforementioned, both in (I) and (II) cases. Ideally, the parameters $\eta$ and $r$ should be chosen accordingly to \eqref{eq:thresholdvalue} and \eqref{eq:r_choice}; however, these bounds are unknown in a real-life experiment, for the radioactive source is generally unknown (and, consequently, so are $E_{min}$ and $E_{max}$).
Both on simulations and real data validations, we found out that taking the parameter $\eta =3 \sigma $, and setting the parameter $r$ so that $\| \mathbf{y} - \mathbf{A} \widehat{\boldsymbol\beta}(r) \|_2 \le \sigma \sqrt{N}$ provided a good compromise between sparsity and good $\ell_2$ precision. It is noticeable to this value of $\eta$ has in our simulation the same order of magnitude as the bound provided in~\eqref{eq:thresholdvalue}, and is only a choice among others.

\subsubsection{Simulation results and discussion}

\begin{figure}[ht]
\includegraphics[width=0.95\linewidth]{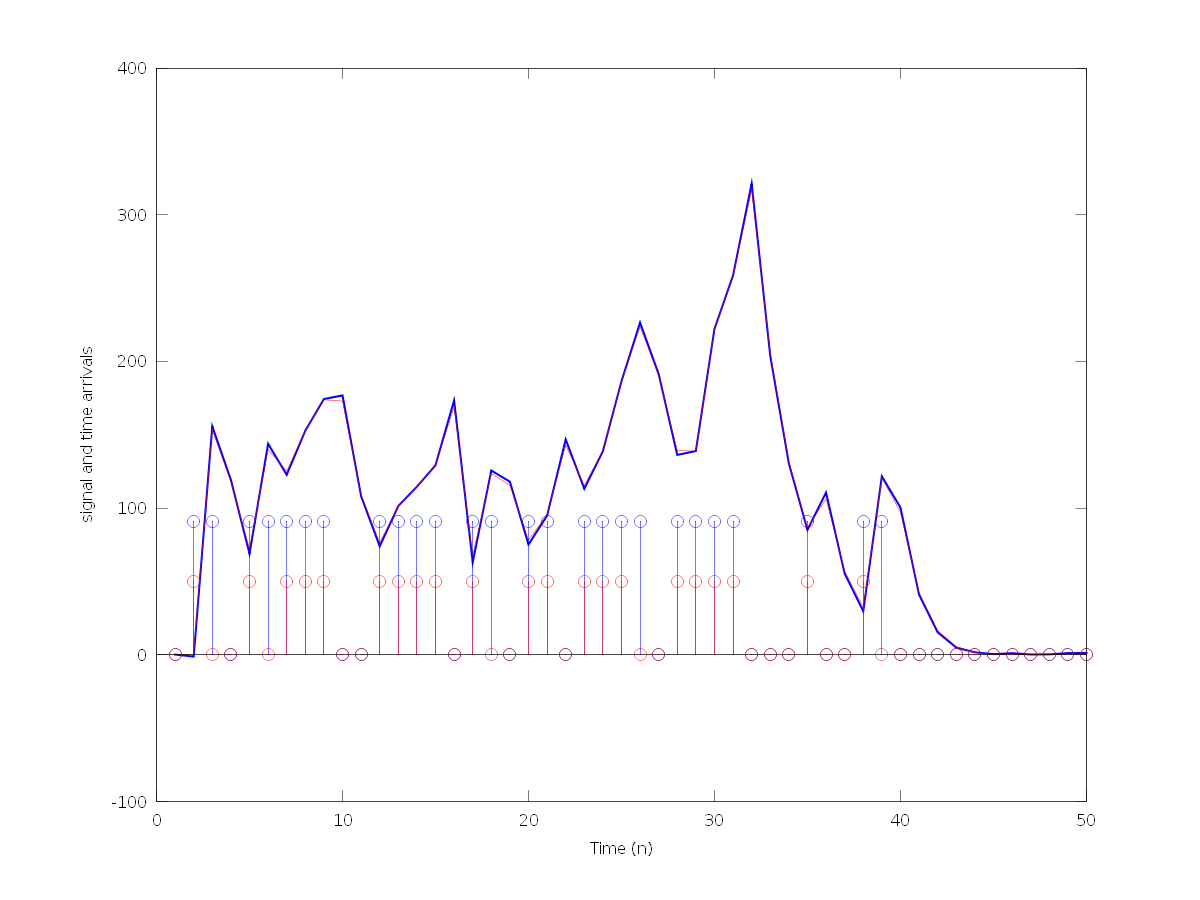}
\caption{Simulated signal (blue) and NNLASSO regressor (red), with associated time arrivals.}
\label{fig:estimation_signal}
\end{figure}

\begin{figure*}[!ht]
     \centering
     \subfigure[Estimated versus actual values of $\lambda$ for the standard method - case (I)]{\label{fig:standard_reallambda_from_dico} \includegraphics[width=.45\textwidth]{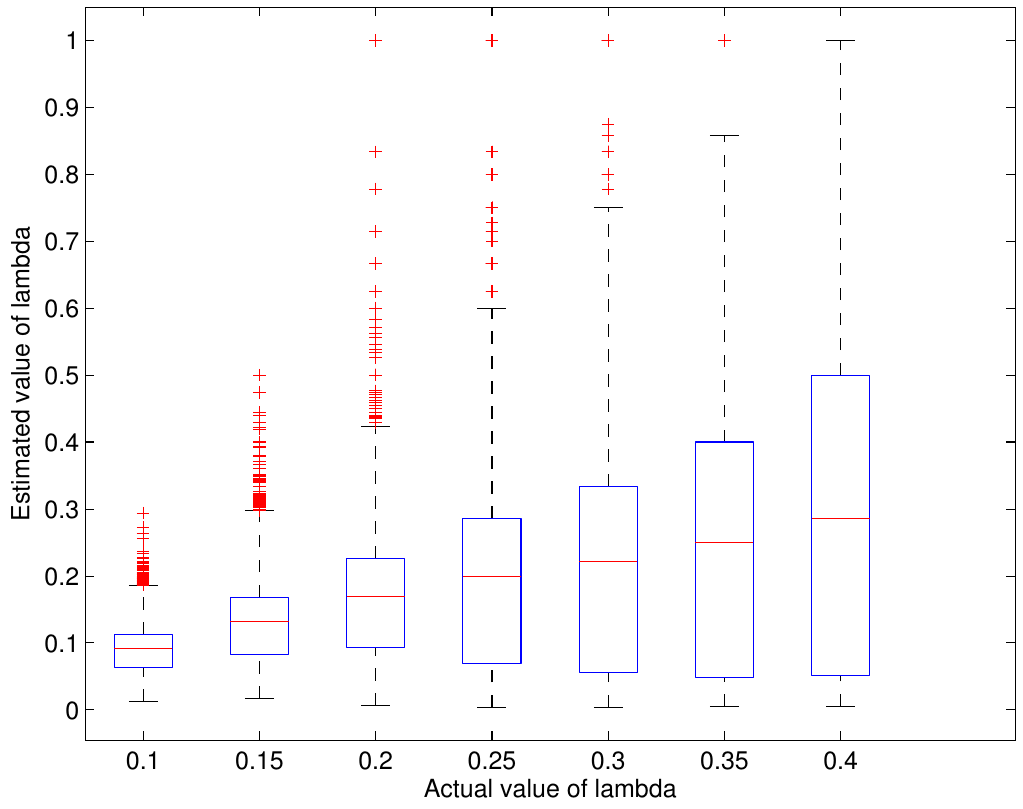}}
     \hspace{.3in}
     \subfigure[Estimated versus actual values of $\lambda$ for the standard method - case (II)]{\label{fig:standard_reallambda_not_from_dico} \includegraphics[width=.45\textwidth]{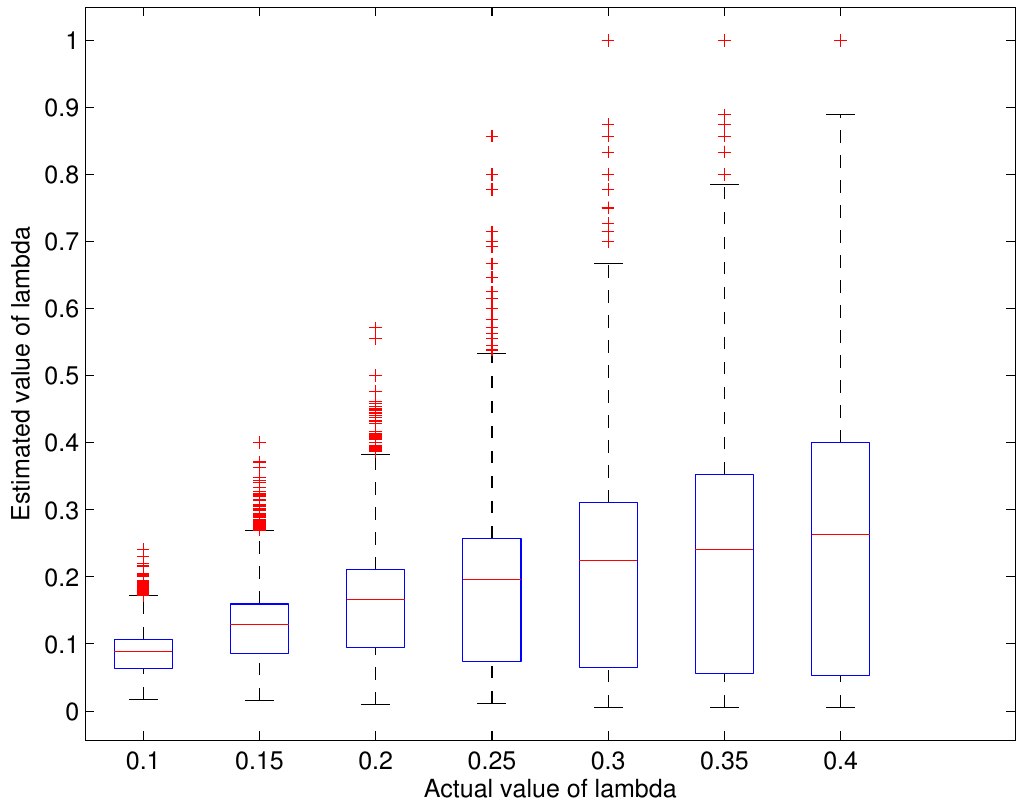}}\\
     \subfigure[Estimated versus actual values of $\lambda$ for the proposed method - case (I)]{\label{fig:Lasso_reallambda_from_dico} \includegraphics[width=.45\textwidth]{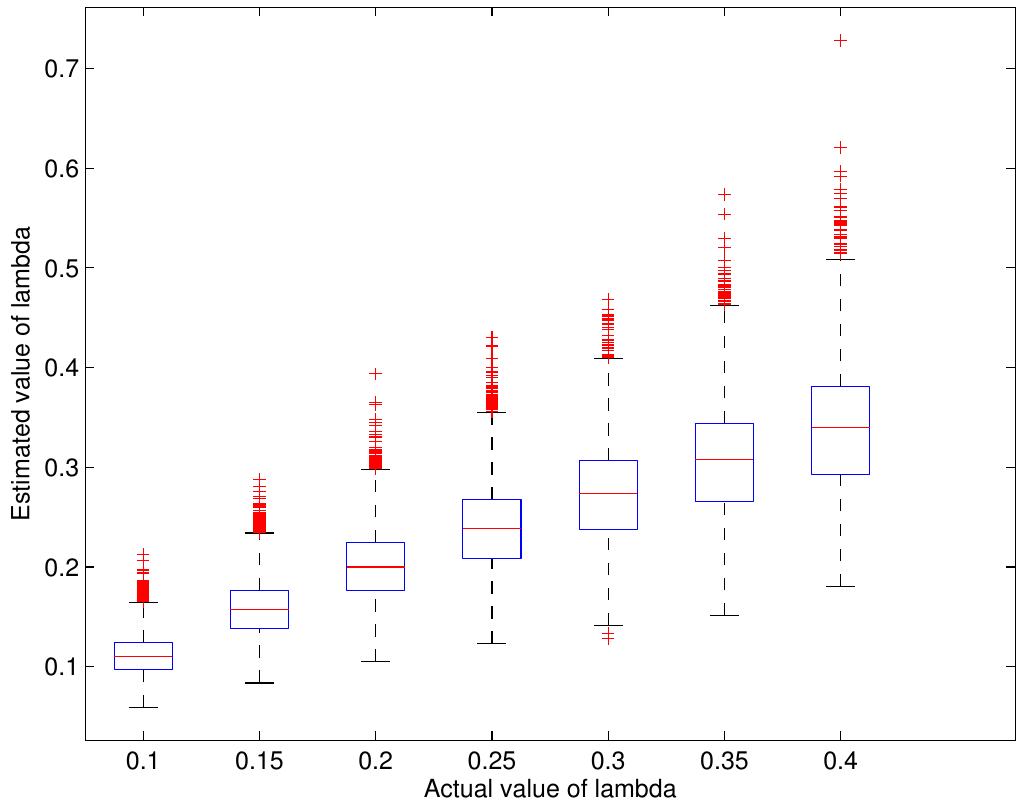}}
     \hspace{.1in}
     \subfigure[Estimated versus actual values of $\lambda$ for the proposed method - case (II)]{\label{fig:Lasso_reallambda_not_from_dico} \includegraphics[width=.45\textwidth]{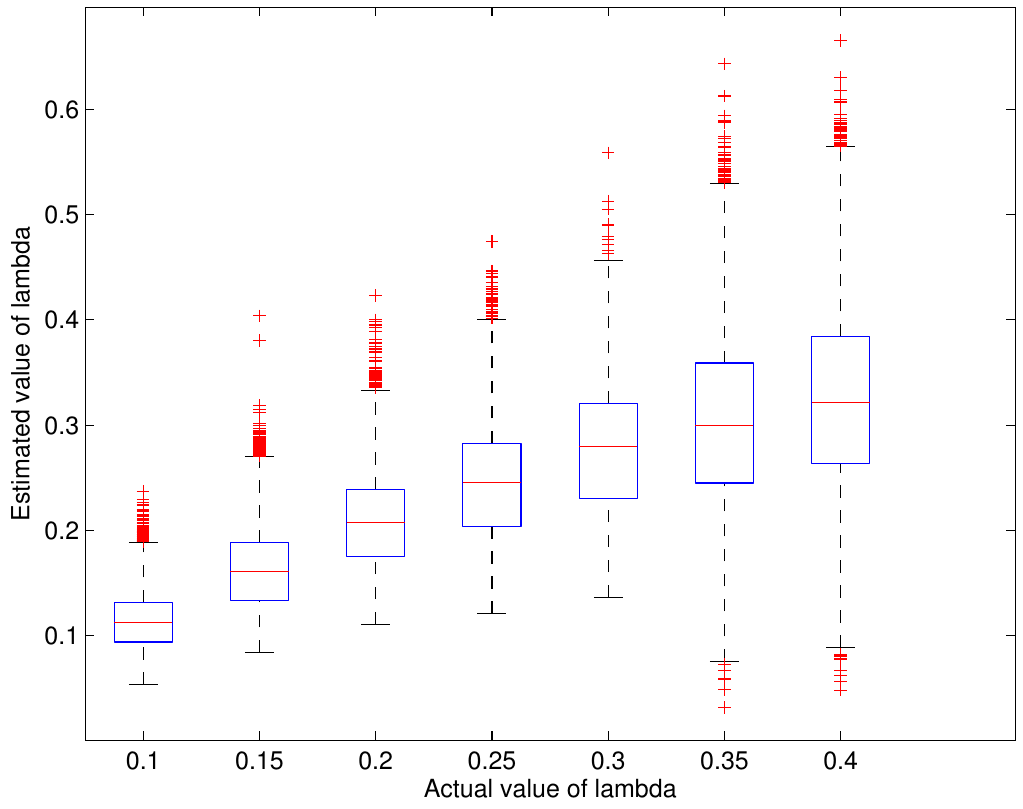}}
     \caption{Distribution of the obtained counting rate estimators ($\eta =3 \sigma $, $r$ is chosen so that $\| \mathbf{y} - \mathbf{A} \widehat{\boldsymbol\beta}(r) \|_2 \le \sigma \sqrt{N}$).}
     \label{fig:results_regression}
\end{figure*}

Figure~\ref{fig:estimation_signal} represents a portion of the simulated signal in case (II) for $\lambda = 1$, as well as the provided estimation and estimated time arrivals. We can observe that the obtained regressor fits well the incoming signal, and that a careful choice of $r_N$ allows to find most of the arrival times.
The boxplots displayed in Figures~\ref{fig:standard_reallambda_from_dico} and~\ref{fig:standard_reallambda_not_from_dico} represent the distribution of the estimators of $\lambda$ (the actual value of $\lambda$ is displayed in the $x$-axis) when using the standard method counting rate estimation, and the results obtained by our method are given in Figures~\ref{fig:Lasso_reallambda_from_dico} and~\ref{fig:Lasso_reallambda_not_from_dico}. It can be seen from these results that the proposed algorithm provides an estimator with smaller variance, thus making it more appropriate for counting rate estimation. 

The high variance in the standard thresholding method can be easily explained. As $\lambda$ increases, so does the number of pileups, hence the number of individual pulses and arrival times are underestimated. Both phenomena yield a poor estimate of $\lambda$. Regarding the estimator obtained with the proposed algorithm, the results obtained in cases where $ \lambda $ is high (e.g. greater than $0.15$) are much better than those of the standard method: we observe a much smaller variance, and for the intensities $0.1$ to $0.2$ the obtained results are very close to the actual counting rate. 

\begin{figure*}[!ht]
     \centering
     \subfigure[Proposed estimate of $\lambda$ versus $\lambda_{opt}$ - case (I)]{\label{fig:Lasso_bestlambda_from_dico} \includegraphics[width=.45\textwidth]{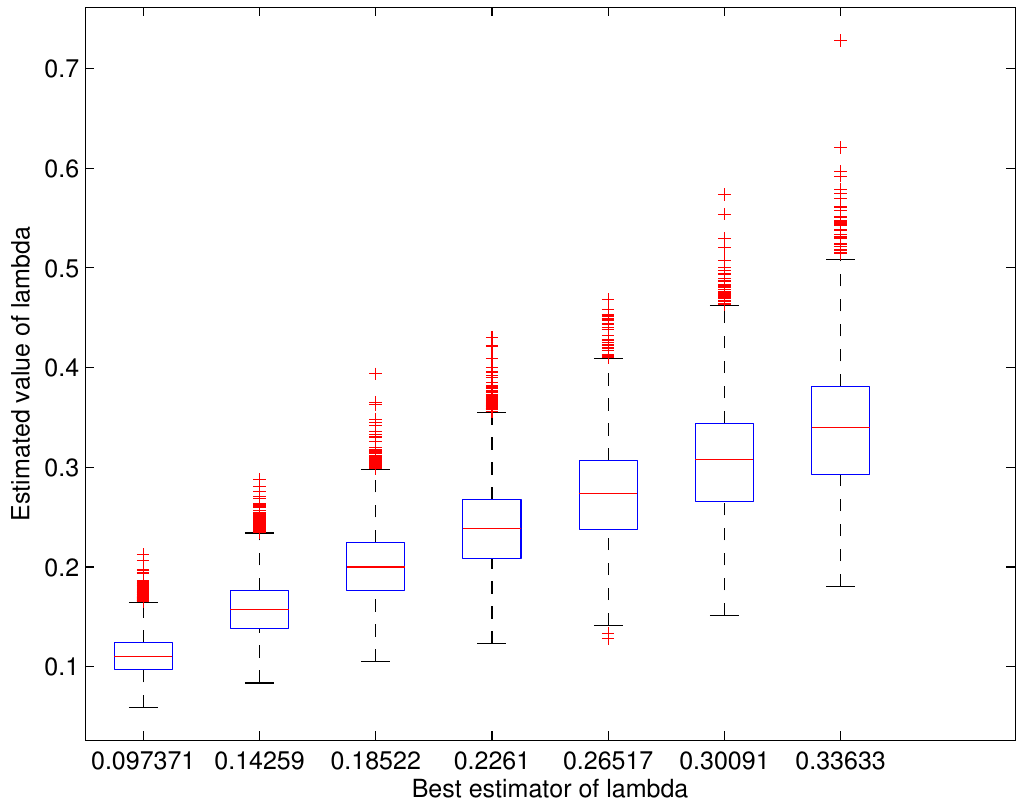}}
     \hspace{.3in}
     \subfigure[Proposed estimate of $\lambda$ versus $\lambda_{opt}$ - case (II)]{\label{fig:Lasso_bestlambda_not_from_dico} \includegraphics[width=.45\textwidth]{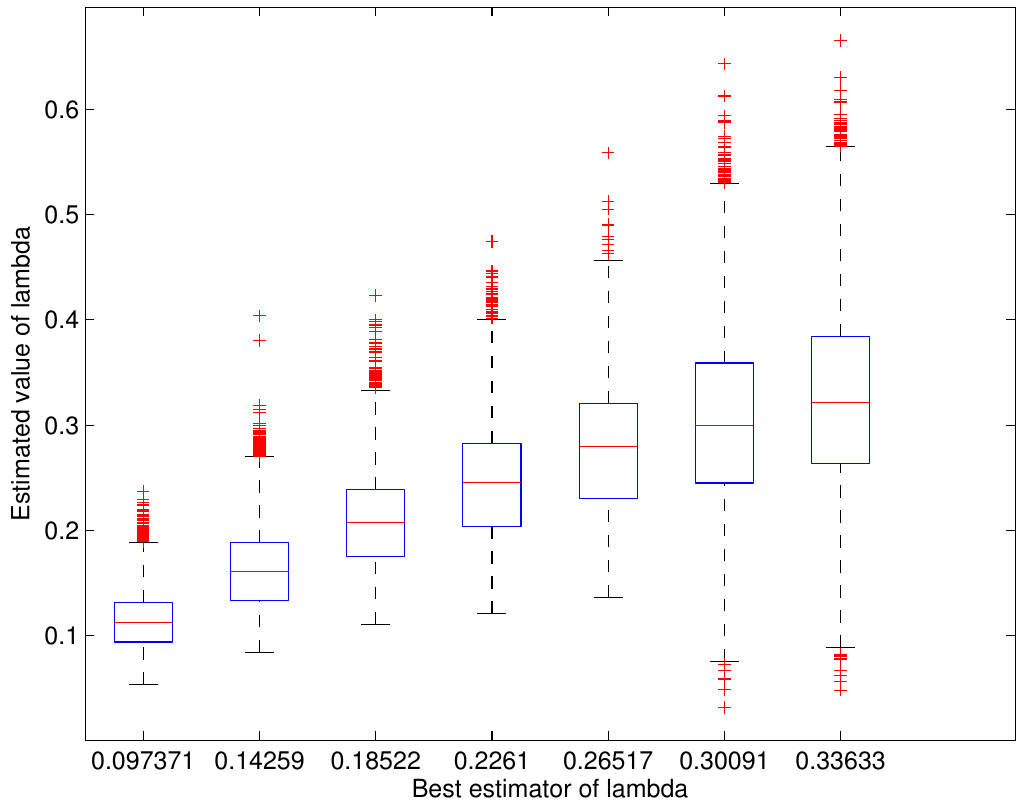}}
     \caption{Comparison of $\hat\lambda$ with $\lambda_{opt}$  ($\eta =3 \sigma $, $r$ is chosen so that $\| \mathbf{y} - \mathbf{A} \widehat{\boldsymbol\beta}(r) \|_2 \le \sigma \sqrt{N}$).}
     \label{fig:results_regression_optimal}
\end{figure*}

When $\lambda $ becomes higher, several pulses are likely to start between two consecutive sampling points. Thus, the suggested algorithm may be misled in treating both as one single impulse, which explains why $\lambda$ is underestimated. However the data is obtained from a sampled signal, therefore the actual $\lambda$ cannot be well estimated when $\lambda \Delta t$ is too high. Indeed, a better insight is obtained when comparing the values of our estimate with $\lambda_{opt}$ instead of $\lambda$.
%
This is done in Figures~\ref{fig:Lasso_bestlambda_from_dico} and~\ref{fig:Lasso_bestlambda_not_from_dico}. We observe an almost linear fit between both estimators, in accordance with Theorems~\ref{th:bounding_intensity_lambdaP} and~\ref{cor:relation_lambdaopt_lambdapost}, thus showing numerically that the proposed approach provides values similar to $\lambda_{opt}$, which is the best estimate we could build from a full knowledge of $T_n$ and of the sampled signal.

\subsection{Applications on real data}

\begin{figure}[!ht]
\includegraphics[width=0.85\linewidth]{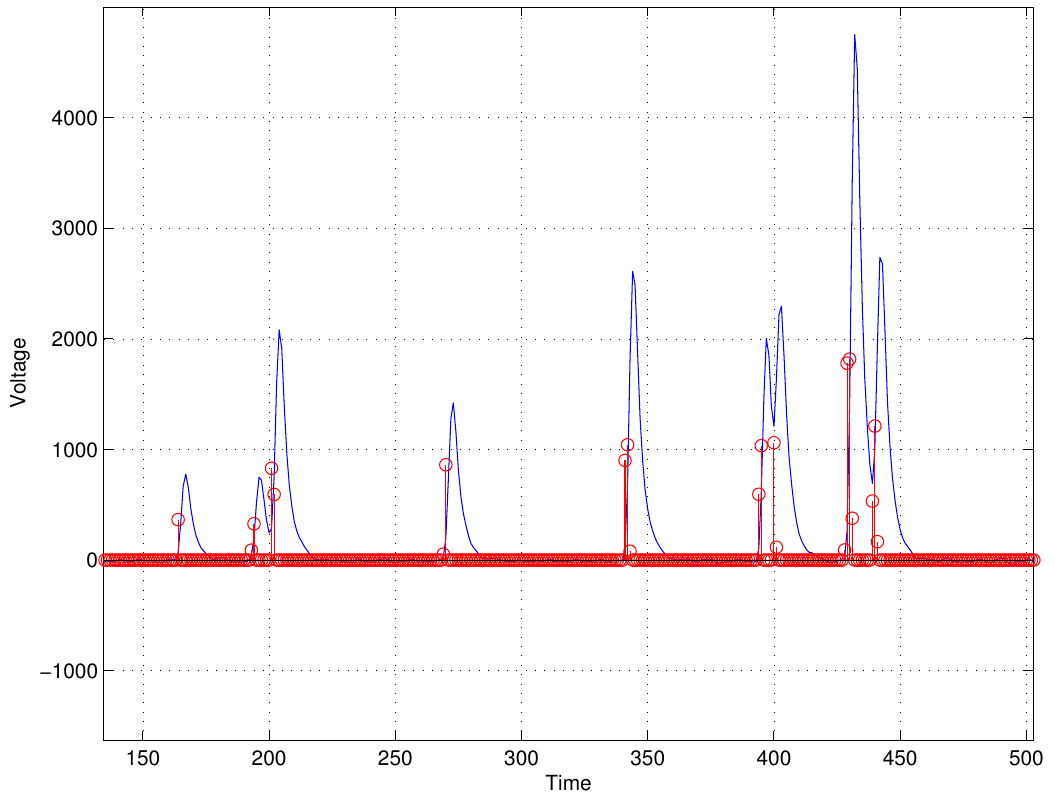}
\caption{Results on real data: input discrete signal (blue), and active/inactive blocks (red). The algorithm parameters are chosen accordingly to the simulation. We observe several well-separated pileups.}
\label{fig:real_signal_lasso}
\end{figure}

We applied the proposed method for counting rate estimation on real spectrometric data from the ADONIS system described in~\cite{adonis:2006}, which is sampled to $10$ MHz. The actual activity is $400000$ photons per second, which corresponds to an intermediate activity. Figure~\ref{fig:real_signal_lasso} shows the use of the proposed algorithm on a real dataset.
It can be observed from the latter figures that a very incomplete dictionary is more than sufficient to retrieve the starting points of each individual pulses. However, the post-processing step we suggest in this paper is required to estimate the activity of the radioactive source. The obtained estimated activity is $3.99\,.10^4$, which conforms both to the simulations and the knowledge of the dataset.

\begin{figure*}[!ht]
     \centering
     \subfigure[Spectrometric signal]{\label{fig:signal_raw} \includegraphics[width=.45\textwidth]{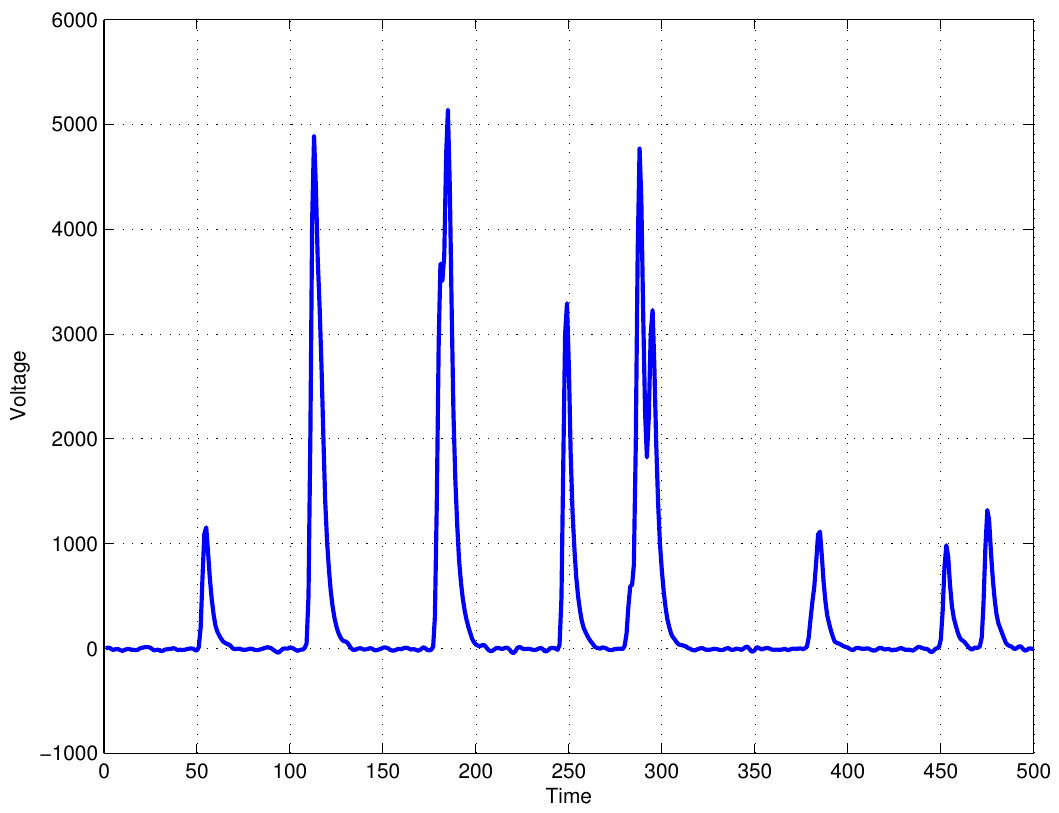}}
     \hspace{.3in}
     \subfigure[$\widehat{\boldsymbol\beta}(r)$ obtained with NNLASSO (parameters as in simulation)]{\label{fig:LASSO_values} \includegraphics[width=.45\textwidth]{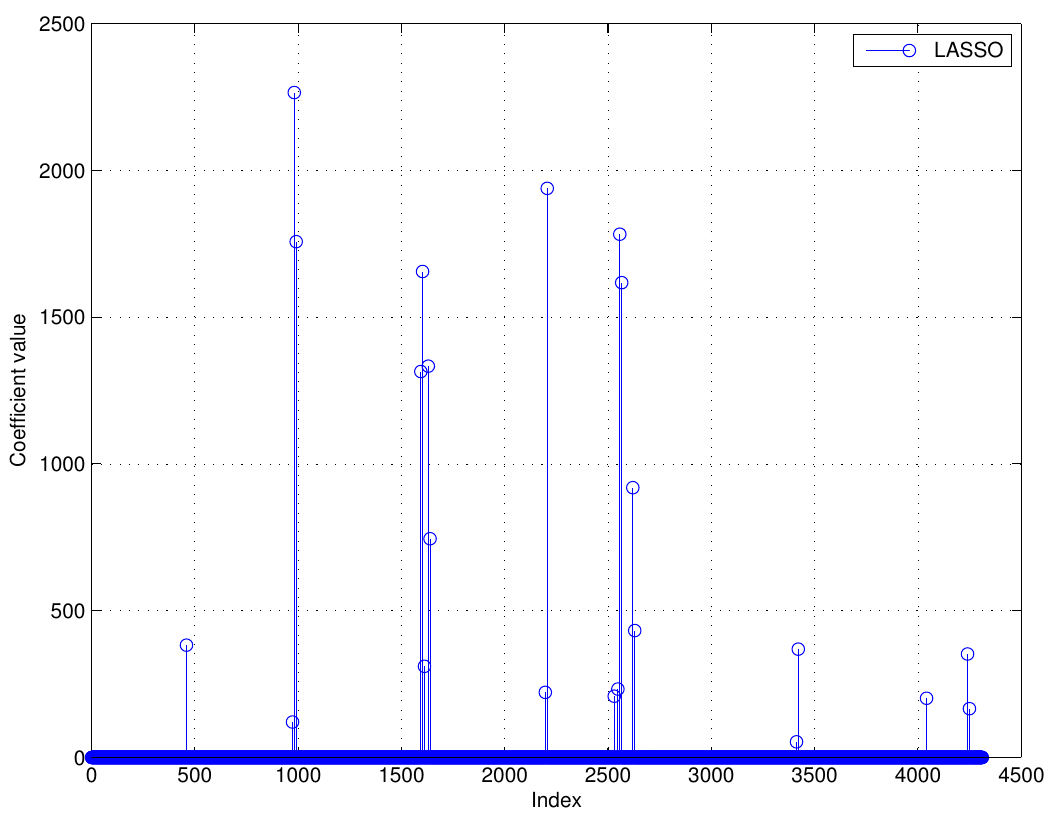}}\\
     \subfigure[$\widehat{\boldsymbol\beta}(r)$ obtained with SBL~\cite{wipf_iterative_2010} (tuning parameter chosen to $7000$)]{\label{fig:SBL_values} \includegraphics[width=.45\textwidth]{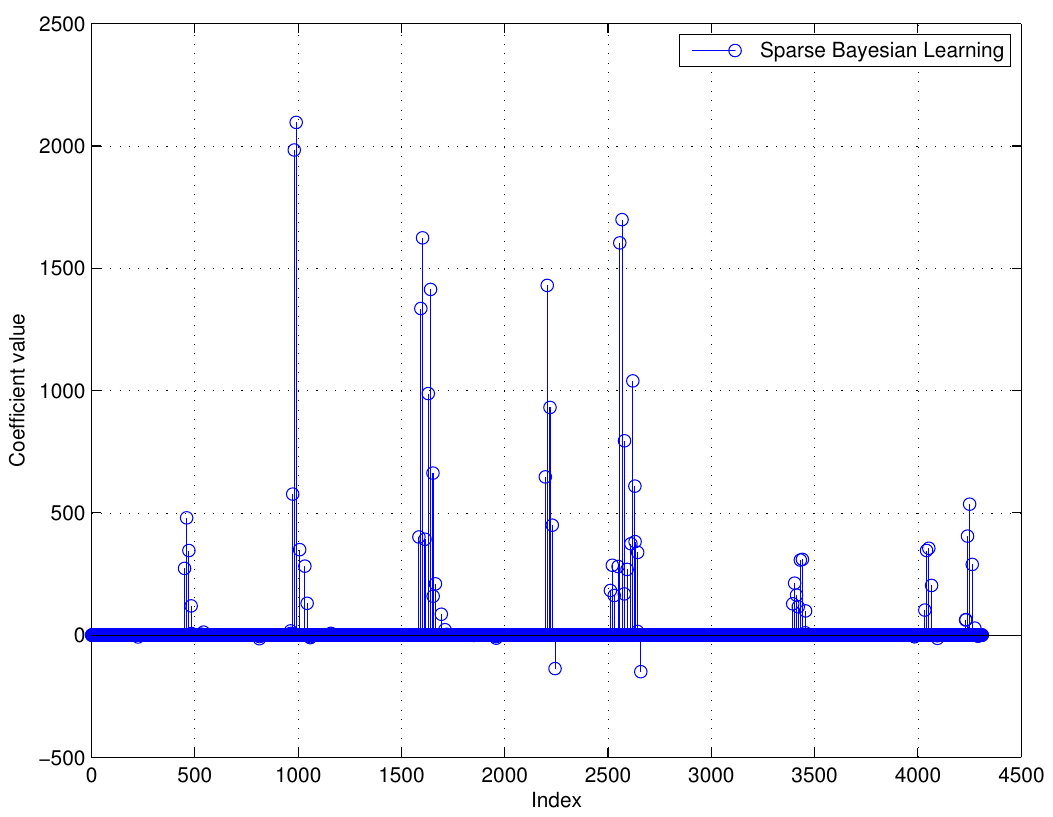}}
     \hspace{.1in}
     \subfigure[$\widehat{\boldsymbol\beta}(r)$ obtained with Reweighted Basis Pursuit Denoising~\cite{candes_enhancing_2008} (10 iterations)]{\label{fig:RWBPDN_values} \includegraphics[width=.45\textwidth]{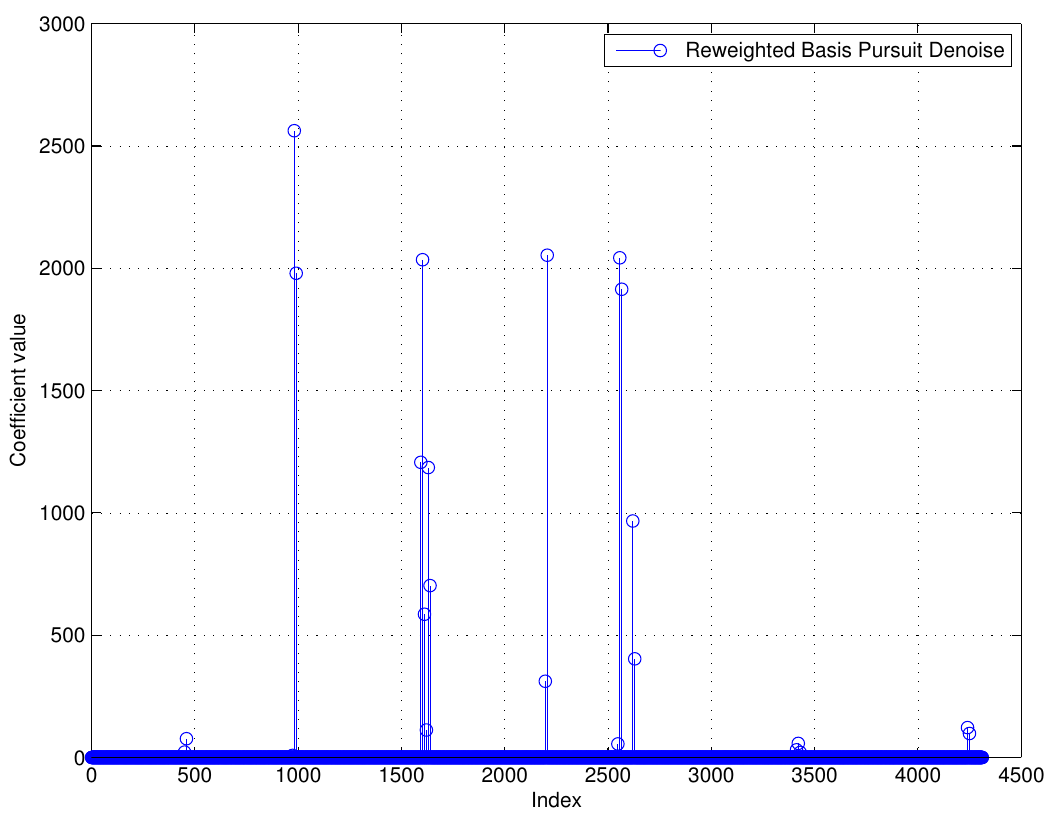}}
     \caption{Comparison of different sparse regression algorithms on real data}
     \label{fig:other_sparse}
\end{figure*}

We illustrate the importance of the post-processing steps for real data in Figure~\ref{fig:other_sparse}, and compare the performances of NNLASSO, the sparse Bayesian learning (SBL) of~\cite{wipf_iterative_2010} and the reweighted $\ell_1$ procedure described in~\cite{candes_enhancing_2008}. We observe that SBL seems to provide a better ``inner-block sparsity'', in the sense that SBL provides active blocks with fewer active coefficients. This is due to the fact that SBL performs usually better than NNLASSO when the columns of $\mathbf{A}$ are highly correlated. However, as it can be seen from the repartition of the coefficients, particularly for the coefficients indexes ranging from $1000$ to $3000$, singles pulses are always reconstructed by means of several active consecutive blocks of $\mathbf{A}$, even with methods providing sparser solutions than NNLASSO. This illustrates that in practice, the post-processing steps cannot be circumvented by improving sparsity. 

\section{Conclusion}

We presented in this paper a method based on sparse representation of a sampled spectrometric signal to estimate the activity of an unknown radioactive source. Based on a crude dictionary, we suggest a post-processed variation of a non-negative LASSO to estimate the number of individual electrical pulses and their arrival times. Results on simulations and real data both emphasize the efficiency of the method, and the small size of the dictionary make the implement for real-time applications accessible. It was theoretically shown that although the standard conditions are not met \emph{per se} for NNLASSO to estimate the actual $P_0$, we can derive some conditions which guarantee that the number of individual pulses and arrival times are well estimated nevertheless. This is made possible by the fact that we do not wish to reconstruct the input signal, but rather find some partial information. Further aspects should focus on the joint estimation of $\lambda$ and the energy distribution, as well as the estimation of the activity in a nonhomogeneous case, and should appear in future contributions. 

\appendices
\label{sec:proofs_appen}

\section{Technical lemmas}
\label{sec:lemmas}

\begin{lem}  \label{lem:bounding_l1normblocks_model}  If  $\ab \bbe \in \mathcal{C}$ then for all index $m \in J(\bbe)$ we have:
\begin{equation*}  \left \| \bbe_{m} \right \|_{1} \leq \frac{E_{\max}}{\min_{i,j} \bg (i,j)^{1/2} }
\end{equation*}
and
\begin{equation*}
\min_{i,j} \bg (i,j)^{1/2} \, \frac{E_{\min}^{2}}{E_{\max}} \leq \left \| \bg \bbe_{m} \right \|_{\infty} 
\end{equation*}
\end{lem}
\begin{pv} For all $m \in J(\bbe )$ we have $ E_{\min}^{2} \leq \bbe_{m}^{T} \bg \bbe_{m} \leq E_{\max}^{2}$, and on the other hand
$$ \min_{i,j} \bg (i,j) \cdot \left \| \bbe_{m} \right \|_{1}^{2} \leq \bbe_{m}^{T} \bg \bbe_{m} \leq E_{\max}^{2} $$
 so the first assertion follows. The second is proved by using the previous result in the following way
\begin{multline*}
E_{\min}^{2} \leq \bbe_{m}^{T} \bg \bbe_{m} \leq \left \| \bbe_{m} \right \|_{1} \left \| \bg \bbe_{m} \right \|_{\infty} \\
\leq  \frac{E_{\max}}{\min_{i,j} \bg (i,j)^{1/2} }   \left \| \bg \bbe_{m} \right \|_{\infty},
\end{multline*}
This concludes the proof.
 \end{pv}

%

\begin{lem} 
\label{lem:proba_consecutive_instants_Poissonian_process}
Suppose a homogeneous Poisson process of intensity $\lambda$ is observed in the interval $[0,T]$, and let $\delta >0$ such that $\lambda^2 T \delta < 1$. The probability that all interarrival times are greater than $\delta $ is bounded from below by $1-\lambda^2 T \delta$. 
\end{lem}
     
\begin{pv}
We compute the probability that one interarrival time is smaller than $\delta$. Denote by $T_n$ the $n$-th point of the process sample path, and by $N_t$ the number of points on $[0,t]$. It is known (see e.g. \cite{baccelli_elements_2002}) that
\begin{multline*}
f_{((T_{1}, \cdots , T_{n} ) \, | N_{T} = n )} (u_1 , \cdots , u_{n} ) \\ = 
  f_{(U_{(1)},U_{(2)},...,U_{(n)})}(u_1,u_2,...,u_n) = \frac{n!}{T^{n}} \, \mathbf{1}_{0=u_{0} \leq u_1 \leq \cdots \leq u_n \leq T}, 
\end{multline*}
where $\{U_{(i)},\ i=1\ldots n\}$ are the order statistics of $n$ independent random variables uniformly distributed on $[0,T]$. 
We get for $n \geq 2$ and $2 \leq i \leq n $ that
 $$ P (  T_{i} - T_{i-1} \leq \delta  | N_{T} = n ) = \frac{n !}{T^{n}} \, \text{Vol} ( \Omega_{i} )  $$
 where $\Omega_{i} \eqdef \{ 0 \leq u_1 \leq \cdots \leq u_n \leq T \, ; \, \, u_i - u_{i-1} \leq \delta \}$ and Vol denotes the volume of the latter space. For all $1 \leq k \leq n $ we set $\text{incr}_k = u_k - u_{k-1} $ (and $u_0 =0 $), so it is equivalent to write
 \begin{multline*}
  \Omega_{i} = \{ \text{incr}_{k} \geq 0, \, 1\leq k \leq n ; \\ \, \, \text{incr}_i \leq \delta \, \text{and} \, \sum_{k=1}^n \text{incr}_k \leq T \}
 \end{multline*} 
  We have now the decomposition of $\Omega_{i}$ along the (disjoint) slices defined by $\text{incr}_i = t , \, \, 0 \leq t \leq \delta $:
 \begin{multline*}
 \Omega_{i} = \bigcup_{0 \leq t \leq \delta } \tilde{\Omega}_{i}(t) ;\\
 \tilde{\Omega}_{i}(t) \eqdef 
 \{ \text{incr}_{j} \geq 0, \, \, \text{for all} \, j \neq i , \, \text{incr}_i = t ; \sum_{j \neq i} \text{incr}_{j} \leq T - t \}
 \end{multline*}
  Integrating now along the variable $t$ we obtain:
 $$ \begin{array}{lllcc}
  \displaystyle \text{Vol} ( \Omega_{i} ) & \displaystyle = \int_{0}^{\delta} \text{Vol} (\tilde{\Omega}_{i}(t)) \, dt \\
    & \displaystyle = \int_{0}^{\delta} \frac{(T-t)^{n-1}}{(n-1)!} \, dt = \frac{1}{n!} \, \left [ T^n - (T- \delta)^n \right ] 
    \end{array} $$
 Hence we have
 $P ( \{ T_{i} - T_{i-1} \leq \delta \} \, | N_{t} = n ) = 1 - \left ( 1 - \frac{\delta}{T} \right )^n $ therefore we get for all $n \geq 2$ that
 \begin{multline*}
 P( T_i - T_{i-1} \leq \delta \, \, \text{for} \, \, \text{some} \, \, 2 \leq i \leq n \, \, | \, N_t = n ) \\ \leq
 (n-1) \, \left [ 1 - \left ( 1 - \frac{\delta}{T} \right )^n \right ]  
 \end{multline*}
 We can of course write that the same probability is equal to $0$ as $n=0,1$. Due to the equality $\displaystyle \sum_{n \geq 2} (n-1) \, \frac{x^n }{n! } = (x-1) \, \exp(x) + 1$, we get by conditioning that the probability that one interarrival time is smaller than $\delta$ is not greater than
 \begin{align*}
&  \exp( - \lambda \, T) \sum_{n \geq 2} \frac{ \lambda^{n} T^{n} }{n! } \, (n-1) \, \left [ 1 - \left ( 1 - \frac{\delta}{T} \right )^n \right ]\\ & = \lambda \, T - \left [ \lambda (T - \delta) - 1 \right ] \, \exp ( - \lambda \, \delta ) - 1
  \end{align*}
The lemma follows from Taylor inequality.
\end{pv}

\section{Proof of Proposition~\ref{prop:first_question}}
\label{sec:proof_q1}

 In its very essence, the proof follows~\cite{wainwright:2009} with mild modifications. We introduce $\widehat{\boldsymbol{\be}}_{0}(r)$ as the minimizer of the NNLASSO problem with sparsity parameter $r>0$ under the additional constraint $J(\widehat{\boldsymbol{\be}}_{0}(r)) \subseteq P_{0}$. The aim is to prove that, under the conditions stated in the result, this vector is a global minimizer of the unconstrained problem. Due to \eqref{eq:estimator_general_def}, the Karush-Kuhn-Tucker (KKT) conditions with the additional constraint are:
\begin{align} \label{eq:kktprop1}
  \frac{\textbf{A}_{P_{0}}^{T}}{N} \left [ \textbf{A}_{P_{0}} \, \boldsymbol{\beta} + \boldsymbol\delta+\boldsymbol\varepsilon
-  \textbf{A}_{P_{0}} \, \widehat{\boldsymbol{\beta}}_{0} (r) \right ] & = r \, \textbf{z}_{P_{0}} \\
  \textbf{z}_{P_{0}} \in (- \infty \, , \, 1 ]^{p |P_{0} |} & \nonumber 
\end{align}
 We deduce from \eqref{eq:kktprop1} the equality 
$\bbe - \widehat{\bbe}_{0} (r) = \bg_{\po , \po}^{-1} \bigl[ r \, \textbf{z}_{\po} - \frac1N \ab_{\po}^{T} \bzeta \bigr ]$. Using this equality, $\widehat{\bbe}_{0} (r)$ will be a global minimizer of NNLASSO  as soon as
\begin{equation*}
   \frac{\textbf{A}_{P_{0}}^{T}}{N} \Bigl [ \textbf{A}_{P_{0}} \, \bg_{\po , \po}^{-1} \bigl( r \, \textbf{z}_{\po} - \frac{\ab_{\po}^{T} \, \bzeta}{N} \bigr )  + \boldsymbol\delta + \boldsymbol\varepsilon \Bigr ]   < \ r \, \1_{\overline{\po}} , 
\end{equation*}
or, equivalently, when
\begin{multline}
 \left [ \frac{\ab_{\overline{\po}}}{\sqrt{N}} \, - \,  \frac{\ab_{\po}}{\sqrt{N}} \, \bg_{\po , \po}^{-1} \bg_{\po , \overline{\po}} \right ]^{T} 
\frac{\bzeta}{\sqrt{N}}  \\
< \ r \left [ \1_{\overline{\po}}  \, - \, \bg_{\overline{\po} , \po} \bg_{\po , \po}^{-1} \, \textbf{z}_{\po} \right ] . 
\label{eq:kktprop1part2}
\end{multline}
For convenience, we now define
$$
\mathbf{H}  \eqdef  \frac{\ab_{\overline{\po}}}{\sqrt{N}} \, - \,  \frac{\ab_{\po}}{\sqrt{N}} \, \bg_{\po , \po}^{-1} \bg_{\po , \overline{\po}} ,
$$
and denote by $H_i$ the $i$-th column of $\mathbf{H}$. Note that $\mathbf{H}$ can be rewritten as $$ \mathbf{H} = \left ( I - \left ( \frac{\ab_{\po} }{\sqrt{N}} \right )  \bg_{\po , \po}^{-1}  \left ( \frac{\ab_{\po} }{\sqrt{N}} \right )^T \right )^{T} \frac{1}{\sqrt{N}}  \ab_{\overline{\po}} ,$$
showing that the columns of $\mathbf{H}$ are the projections of the normalized columns of $\ab_{\overline{\po}}$ onto the orthogonal complement of the columns of $\ab_{\po} $. It thus follows that all columns the $H_i$'s have normalized $\ell_{2} $-norm bounded by $1$ since this is true for $\ab_{\overline{\po}}$.  Due to assumption (\ref{eq:block_irrep_cond}), it is sufficient that the following condition holds to get \eqref{eq:kktprop1part2}:
\begin{equation} \label{eq:sufficient_for_KKT}
   \max_{i} \frac{1}{\sqrt{N}} H_i^T \left ( \bdelta + \beps \right )  < r \, \eta_{0} \ ; 
\end{equation}
we now use the fact that the random variable $H_{i}^T \beps $ is Gaussian of variance less than $\sigma^{2}$, consequently:
\begin{align}
& P \left( \max_{i} \frac{1}{\sqrt{N}} H_{i}^T \beps \geq \frac{r \, \eta_{0} }{2} \right)  \nonumber \\
& \leq \sum_{i} P \left (H_{i}^T \beps  \geq  \frac{\sqrt{N} \, r \, \eta_{0} }{2} \right ) \nonumber \\
& \leq  p (N- | P |) \,  \mathfrak{t} \left ( \frac{\sqrt{N} \, r \, \eta_{0} }{ 2 \, \sigma} \right )  \label{eq:sufficient_cond_minimum}
\end{align}
In order to make (\ref{eq:sufficient_cond_minimum}) tend to $0$, we need that 
 $$
    r \geq \frac{2 \, \sqrt{2} \, \sigma }{\eta_{0}} \, \sqrt{\frac{\log (N - | P |)\,  p }{N}} \ .
$$
Now we have also $\frac{1}{\sqrt{N}} H_i^{T} \, \bdelta  \leq \alpha $ by Cauchy-Schwarz inequality, and we have $ \alpha < r \eta_{0} / 2$ as soon as $ r > 2 \alpha / \eta_{0} $. The result follows. 

\section{Proof of Proposition \ref{prop:bounds_sparsitylevel_LASSOregressor}}
\label{proof:bounds_sparsitylevel_LASSOregressor}

We keep the same notations as in Appendix~\ref{sec:proof_q1}. Let $r>0$ be any sparsity parameter, and $\widehat{\bbe}(r)$ be the corresponding NNLASSO regressor, denoted shortly by $\widehat{\bbe}$ in the rest of this proof. The KKT conditions, combined with the inequality derived from Cauchy-Schwarz $ \frac{1}{N} \ab^{T} \bdelta \leq \alpha \1_{p N}$, yields
\begin{equation*}
\frac{\textbf{A}^{T}}{N} \left [ \ab (\bbe - \widehat{\bbe}) + \boldsymbol\delta+\boldsymbol\varepsilon \right] \leq r \1_{p N},  
\end{equation*}
or, equivalently,
\begin{equation}
\frac{\ab^{T} \ab}{N} \bbe - (r + \alpha) \1_{p N} + \frac{\ab^{T} \beps}{N} \leq  \frac{\ab^{T} \ab}{N} \widehat{\bbe} .\label{eq:kkt_gene}
\end{equation}
For some $k$ in $P_0$, it follows from~\eqref{eq:kkt_gene} that
$$  \| \textbf{G} \bbe_{k} \|_{\infty} - (r + \alpha) - \frac1N  \| \ab_{k}^{T} \beps \|_{\infty} \leq 
 \| \bg_{k, V_{\tau}(k)} \widehat{\bbe}_{V_{\tau}(k)}  \|_{\infty} .$$
Define $0 \le \rho \le 1$; since all the coefficients of the Gram matrices considered are bounded by $1$, we 
have
%
\begin{multline}
   \| \textbf{G} \bbe_{k} \|_{\infty} - (r + \alpha) - \frac1N \| \ab_{k}^{T} \beps \|_{\infty} \\ \leq  (1- \rho) \, \sum_{l \in \mathcal{T}_{\rho}(k)} \| \widehat{\bbe}_{l}  \|_{1} + \rho  \| \widehat{\bbe}_{V_{\tau}(k)}  \|_{1} \ . \label{eq:ineqrho}
\end{multline}  
We can say now two things: first, when $r +\alpha <  \left \| \textbf{G} \bbe_{k} \right \|_{\infty}  $, we have 
\begin{multline*}
\Pr \left ( \frac1N  \| \ab_{k}^{T} \beps \|_{\infty} >  \frac{\left \| \textbf{G} \bbe_{k} \right \|_{\infty} - (r + \alpha) }{2} \right ) \\ < p \mathfrak{t} \left ( \sqrt{N} \, \frac{\left \| \textbf{G} \bbe_{k} \right \|_{\infty} - (r + \alpha) }{2 \sigma} \right ) ; 
\end{multline*}
secondly, for any column $A_n$ of $\ab $ which is active in $\wbbe$, \eqref{eq:kkt_gene} reduces to the equality $\frac{A_n^T}{N} \left ( \ab \bbe + \bdelta + \beps \right ) - r =  \frac{A_n^T}{N}  \ab \widehat{\bbe} $, and since $\wbbe$ has non negative entries it follows that $\widehat\beta_n$ is smaller than $\frac1N A_n^T \left ( \ab \bbe + \bdelta + \beps \right ) - r $; now summing up these inequalities overall such $n \in  Supp(\wbbe_{ V_{\tau}(k)})$ yields 
\begin{multline} \label{eq:sum_n}
  \| \wbbe_{V_{\tau}(k)} \|_{1} \leq \frac1N  \| \ab^{T}_{V_{\tau}(k) } \ab \bbe \|_{1} + \\
\sum_{ n \in Supp(\wbbe_{ V_{\tau}(k)})} \bigl ( \frac1N A_n^T \beps - r + \alpha \bigr ) .
\end{multline}
 Since every $n \in Supp(\wbbe_{ V_{\tau}(k)}) $ can be expressed as $n= p  l + s $ with $l$ in $V_{\tau}(k) \cap J(\widehat{\bbe})$ and $s \leq p$, we have
\begin{multline*}
\Pr \Bigl (  \sum_{n \in Supp(\wbbe_{ V_{\tau}(k)})} \bigl( \frac1N A_n^T \beps - r + \alpha \bigr) > 0  \Bigr)  \\ \leq p (2 \tau +1 ) \mathfrak{t} \Bigl ( \sqrt{N}  \, \frac{r - \alpha}{\sigma} \Bigr ),
\end{multline*} 
thus \eqref{eq:sum_n} yields
\begin{multline}
\label{eq:pr} 
\Pr \Bigl(  \| \wbbe_{V_{\tau}(k)}  \|_{1} \leq \frac1N  \| \ab^{T}_{V_{\tau}(k) } \ab \bbe \|_{1} \Bigr ) \\
> 1 -  p (2 \tau +1 ) \mathfrak{t} \Bigl ( \sqrt{N}  \, \frac{r - \alpha}{\sigma} \Bigr ) .
\end{multline}
 So far, we showed that 
\begin{multline}
 \label{eq:detect_bloc_l1}
 \frac{1}{(1- \rho) | \mathcal{T}_{\rho}(k) |} \Bigl[  \frac{\left \| \textbf{G} \bbe_{k} \right \|_{\infty} - (r + \alpha) }{2}  -   \rho  \frac1N  \| \ab^{T}_{V_{\tau}(k) } \ab \bbe \|_{1}  \Bigr ] \\
\leq \max_{l \in \mathcal{T}_{\rho}(k)}
\| \wbbe_{l} \|_{1} 
\end{multline}
with probability greater than 
$$
1 -  p \, \mathfrak{t} \Bigl ( \sqrt{N} \, \frac{\| \textbf{G} \bbe_{k}  \|_{\infty} - (r + \alpha) }{2 \sigma} \Bigr ) -  p (2 \tau +1 ) \mathfrak{t} \Bigl ( \sqrt{N}  \, \frac{r - \alpha}{\sigma} \Bigr ) .
$$
Now, choosing $r$ accordingly to \eqref{eq:r_choice} and using Lemma~\ref{lem:bounding_l1normblocks_model} yields
$$\frac{\left \| \textbf{G} \bbe_{k} \right \|_{\infty} - (r + \alpha) }{2} >  \frac{E_{\min}^{2} \, \min_{i,j} \bg(i,j)^{1/2}}{4 \, E_{\max}}, $$
and if $\rho$ is taken equal to $0$ the LHS of \eqref{eq:detect_bloc_l1} is greater than $\eta$ as chosen in~\eqref{eq:thresholdvalue}. On the other hand, the term $\frac1N  \| \ab^{T}_{V_{\tau}(k) }  \ab \bbe \|_{1} $ can be bounded by means of Lemma~\ref{lem:bounding_l1normblocks_model} as follows:
\begin{align*}
&\frac1N  \| \ab^{T}_{V_{\tau}(k) }  \ab \bbe \|_{1}  \leq \sum_{l \in V_{\tau}(k) } \sum_{m \in 
V_{\tau}(\{l\})}  \| \bg_{\{l\},\{m\}} \, \bbe_{m}  \|_{1}  \\
& \leq  p \sum_{l \in V_{\tau}(k) } \sum_{m \in V_{\tau}(\{l\})}  ( \max_{i,j} \bg_{\{l\},\{m\}}(i,j) )  \| \bbe_{m} \|_{1} \\
&  \leq \frac{(2 \tau +1) \mathcal{G} p \, E_{\max}}{ \min_{i,j} \bg (i,j)^{1/2}} .
\end{align*}
Therefore, if~\eqref{eq:r_choice} holds, we introduce
$$
C_{r,\rho} \eqdef \frac{E_{\min}^{2}  \min_{i,j} \bg(i,j)^{1/2}}{E_{max}}  - \alpha - r -  \frac{ (2 \tau +1) \mathcal{G}  p E_{\max}  }{\min_{i,j} \bg(i,j)^{1/2} }  \rho
$$
and
\begin{equation}
\label{eq:convenientrholasso}
 \rho_{r} \eqdef \sup \Bigl \{ \rho \in [0,1] \, \, ; \ \eta \leq \frac{ C_{r,\rho}  }{(1- \rho) | \mathcal{T}_{\rho}(k) |}  \Bigr \} ,
\end{equation}
thus inequality \eqref{eq:detect_bloc_l1} shows that if $k \in P_{0}$ then there exists $m \in \mathcal{T}_{\rho_{r}}(k)$ such that 
\begin{multline*}
\Pr  ( \| \wbbe_{m} \|_{1} \geq \eta  ) > 1 - p  \mathfrak{t} \Bigl ( \sqrt{N} \frac{E_{\min}^{2} \, \min_{i,j} \bg(i,j)^{1/2}}{4 \, E_{\max} \, \sigma} \Bigr ) \\ - (2 \tau +1 )p \mathfrak{t} \Bigl ( \sqrt{N} \, \frac{r - \alpha}{\sigma} \Bigr )  ,
\end{multline*}
which completes the proof of the first part of Proposition~\ref{prop:bounds_sparsitylevel_LASSOregressor}. The converse part is proved using a similar argument as follows. For any $k$, we denote by $\mathrm{Supp}(\wbbe_{k})$ the set of indices of the non-zero entries in the vector $\wbbe_{k}$. For $k$ chosen such that $ \| \wbbe_{k} \|_{1} \geq \eta $, then for all $n$ in $\mathrm{Supp}(\wbbe_{k})$ we have
$\frac1N A_n^T\ab \bbe = r + \frac1N A_n^T \ab \wbbe - \frac1N A_n^T\bzeta $. Considerations analog to those yielding \eqref{eq:sum_n},\eqref{eq:pr} allow to obtain:
\begin{multline*}
\sum_{n \in \mathrm{Supp} (\wbbe_{k})} \frac1N A_n^T \ab \bbe \geq  \sum_{n \in \mathrm{Supp} (\wbbe_{k})}  \Bigl [ r - \alpha - \frac1N A_n^T \beps  \Bigr ] \\
+ \sum_{ n \in \mathrm{Supp} (\wbbe_{k})} \frac1N A_n^T\ab_{k}  \wbbe_{k} 
\end{multline*}
hence
\begin{multline}
\label{eq:minoration}
\Pr \bigg(  \sum_{n \in \mathrm{Supp} (\wbbe_{k})} \frac1N A_n^T \ab \bbe > \min_{i,j} \bg(i,j)  \| \wbbe_{k}  \|_{0} \, \eta \bigg ) \\
> 1 - \| \wbbe_{k}  \|_{0} \, \mathfrak{t} \Bigl( \sqrt{N} \, \frac{r - \alpha}{\sigma} \Bigr ) .
\end{multline}
Now let $\rho$ in $[0,1]$, we can write similarly to \eqref{eq:ineqrho} 
\begin{align}
& \sum_{n \in \mathrm{Supp} (\wbbe_{k})} \frac1N A_n^T \ab \bbe \nonumber \\
& =  \sum_{n \in \mathrm{Supp} (\wbbe_{k})} \frac1N A_n^T \ab 
( \bbe_{\mathcal{T}_{\rho}(k)} +  \bbe_{\overline{\mathcal{T}_{\rho}(k)}} )\nonumber \\
 & <  \| \wbbe_{k} \|_{0} \bigl ( \| \bbe_{\mathcal{T}_{\rho}(k)}  \|_{1} + \rho \|  \bbe_{\overline{\mathcal{T}_{\rho}(k)}
\cap V_{\tau}(k)}  \|_{1} \bigr ) .
\end{align}
Since by Lemma~\ref{lem:bounding_l1normblocks_model} we have $$\|  \bbe_{\overline{\mathcal{T}_{\rho}(k)}
\cap V_{\tau}(k)} \|_{1} \leq \frac{| \overline{\mathcal{T}_{\rho}(k)}
\cap V_{\tau}(k) |  E_{\max}} {\min_{i,j} \bg(i,j))^{1/2}},$$ \eqref{eq:minoration} leads to introduce the next correlation level, similarly to \eqref{eq:convenientrholasso}:
\begin{multline*}
\mu \eqdef  \sup \Bigl\{ \rho \in [0;1] \ ; \ \rho \, |  \overline{\mathcal{T}_{\rho}(k)}
\cap V_{\tau}(k) |  \\ \leq \, \frac{\eta \cdot \min_{i,j} \bg(i,j)^{3/2}}{E_{\max}} \Bigr \} .
\end{multline*}
Thus, \eqref{eq:minoration} implies that whenever $\| \wbbe_{k} \|_{1} \geq \eta $, one has
$$ \Pr  ( P_{0} \cap \mathcal{T}_{\mu}(k) \neq \emptyset  ) > 1 - \| \wbbe_{k}  \|_{0} \, \mathfrak{t} \Bigl( \sqrt{N} \, \frac{r - \alpha}{\sigma} \Bigr ), $$
which completes the proof.

\section{Proofs of Theorem \ref{th:bounding_intensity_lambdaP} and Theorem~\ref{cor:relation_lambdaopt_lambdapost}}
\label{app:bounding_intensity_lambdaP}

We first prove Theorem~\ref{th:bounding_intensity_lambdaP} : by Proposition~\ref{prop:bounds_sparsitylevel_LASSOregressor}, we have
\begin{multline*}
\Pr \left ( \max P_{0} - a_{\rho} \leq \max J(\wbbe(r,\eta)) \right ) > 1 \\ - p \mathfrak{t} \left ( \sqrt{N} \, \frac{E_{\min}^{2} \, \min_{i,j} \bg(i,j)^{1/2}}{4 \, E_{\max} \, \sigma} \right ) \\ - p (2 \tau +1 ) \mathfrak{t} \left ( \sqrt{N}  \, \frac{r - \alpha}{\sigma} \right )
\end{multline*}
and moreover each of the maximal distinct intervals contained in $\bigcup_{1 \leq j \leq \widehat{M}} V_{a_{\mu}} ( \widehat{T}_{j} )  $  intersects $P_{0}$ under probability greater than 
\begin{multline*}
1 - \sum_{k \in J(\wbbe(r,\eta))} \| \wbbe_{k}(r)  \|_{0}  \mathfrak{t} \Bigl( \sqrt{N} \, \frac{r - \alpha}{\sigma} \Bigr ) \\  \geq  1 -  \| \wbbe(r) \|_{0}  \mathfrak{t} \Bigl( \sqrt{N} \, \frac{r - \alpha}{\sigma} \Bigr )
\end{multline*}
thus under the same probability we have $I  (\bigcup_{1 \leq j \leq \widehat{M}} V_{a_{\mu}} ( \widehat{T}_{j} )  ) \leq |P_{0}|$, and combining these two results we obtain the lower bound 
$$
\lambda_{opt} \geq \frac{1}{\Delta t} \frac{I \left ( \bigcup_{1 \leq j \leq \widehat{M}} \mathcal{T}_{\mu} ( \widehat{T}_{j} ) \right )}{a_{\rho} + \max J(\wbbe(r,\eta))} , 
$$
which is equivalent to \eqref{eq:lambdaoptupbound} after factorization by $\widehat{\lambda}(r,\eta)$. 

Theorem~\ref{cor:relation_lambdaopt_lambdapost} is proved in a similar manner. Using again Proposition.~\eqref{prop:bounds_sparsitylevel_LASSOregressor}, we have
\begin{multline*}
\Pr ( \max J(\wbbe(r,\eta)) - a_{\mu} \leq \max P_{0} ) \\ > 1- \| \wbbe_{\max J(\wbbe(r,\eta))} \|_{0}  \mathfrak{t} \Bigl( \sqrt{N} \, \frac{r - \alpha}{\sigma} \Bigr ) .
\end{multline*}
Using the same argument as above, we also get that
$$ | J(\wbbe(r,\eta)) | \geq I \left ( \bigcup_{k \in P_{0}} V_{a_{\rho}}(k) \right ) $$
 with probability greater than 
\begin{multline*}
1 - |P_{0}| \Biggl [ p \mathfrak{t} \left ( \sqrt{N} \, \frac{E_{\min}^{2} \, \min_{i,j} \bg(i,j)^{1/2}}{4 \, E_{\max} \, \sigma} \right )  \\ - p (2 \tau +1 ) \mathfrak{t} \left ( \sqrt{N}  \, \frac{r - \alpha}{\sigma} \right ) \Biggr ]
\end{multline*}
Now the $a_{\rho}-$neighbourhoods $V_{a_{\rho}}(k) $ for $k \in P_{0}$ are all disjoint with probability bounded as in Lemma~\ref{lem:proba_consecutive_instants_Poissonian_process} when taking $T=N\Delta t$; the result follows

\section*{Acknowledgment}
The authors would like to thank the anonymous reviewers whose careful reading and comments greatly helped to improve the readability of the paper.

\bibliographystyle{IEEEtran}
\bibliography{IEEEabrv,Bibliographie_lasso_journal}

%








\begin{IEEEbiography}[{\includegraphics[width=1in,height=1.25in,clip,keepaspectratio]{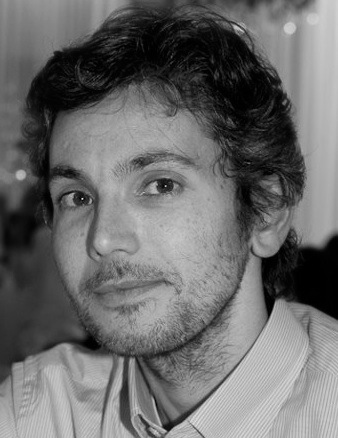}}]{Yann Sepulcre}
Yann Sepulcre was born in Paris in 1977, and received an M.Sc in mathematics from Paris 7 university, with a specialization in algebraic geometry. He received a Ph.D in complex algebraic geometry from the same university in 2004.

He has held a post-doctoral position at Bar-Ilan University in 2004-2006, then started working in applied fields such as computer vision and statistical learning. He is currently a lecturer both in Shamoon college of Engineering and Jerusalem college of Engineering, Israel. His main research interests include statistical signal processing and applications of algebraic geometry in computer vision. 
\end{IEEEbiography}

\begin{IEEEbiography}[{\includegraphics[width=1in,height=1.25in,clip,keepaspectratio]{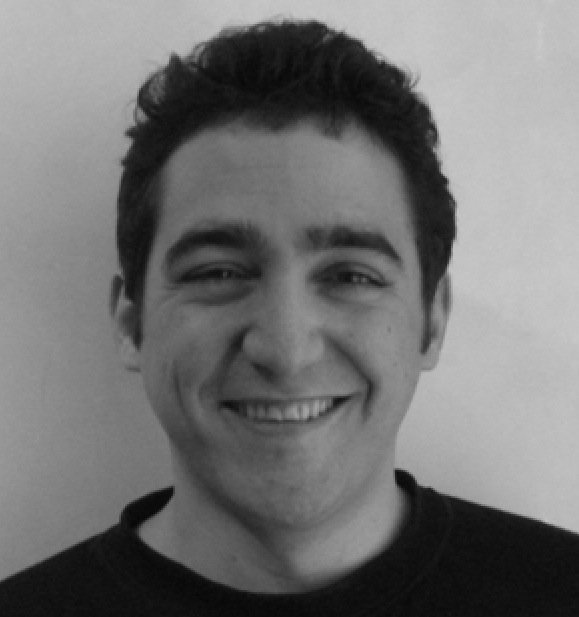}}]{Thomas Trigano}
(M' 10) was born in Paris, France in 1978, and received an M.Sc. in engineering from the T\'el\'ecom Paris Tech (France) and an M.Sc in Applied Probability from Paris VI University (France) in 2001. He recieved the Ph.D. degree in signal processing from the T\'el\'ecom Paris Tech in 2005.

From 2006 to 2008 he received a postdoctoral fellowship from the Hebrew University of Jersualem in the department of statistics. Since 2008 he is senior lecturer in department of Electrical Engineering in Shamoon College of Engineering, Israel. His main research interests include applied statistics, statistical signal processing, pattern recognition and communications. 

\end{IEEEbiography}

\begin{IEEEbiography}[{\includegraphics[width=1in,height=1.25in,clip,keepaspectratio]{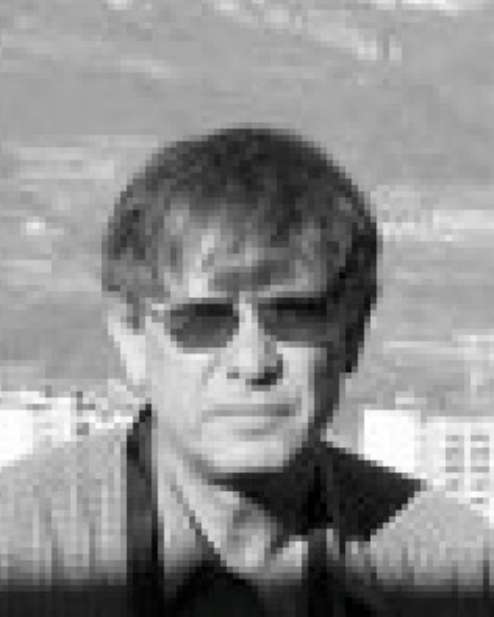}}]{Ya'acov Ritov}
was born in Jerusalem, Israel, in 1951. He received the  B.Sc. and M.Sc. in electrical engineering in the Technion, Haifa, Israel in 1973 and 1980. He received his Ph.D. in statistics from The Hebrew University, Jerusalem, Israel, in 1983.

Since 1984 he has been working in the department of statistics in The Hebrew University of Jerusalem, Israel, and is currently a Full Professor. His research interests include asymptotic theory of estimators for semi-parametric/non-parametric models, statistical analysis of stochastic processes, survival analysis, change detection and signal processing.

Prof. Ritov is a fellow of the Institute of Mathematical Statistics.
\end{IEEEbiography}

\end{document}